\documentclass[sigconf]{acmart}
\acmConference[EASE 2025]{The 29th International Conference on Evaluation and Assessment in Software Engineering}{17–20 June, 2025}{Istanbul, Türkiye}

\usepackage{algorithm2e}
\usepackage{xcolor}

\SetAlFnt{\footnotesize}

\SetAlCapSty{xAlCapSty}

\SetCommentSty{xCommentSty}

\SetNlSty{mynlfont}{}{} 

\LinesNumbered

\SetSideCommentRight

\DontPrintSemicolon

\RestyleAlgo{algoruled}

\usepackage{hyperref}
\usepackage{mdframed}
\usepackage{lipsum}
\usepackage{amsmath}
\usepackage{amsfonts}
\usepackage{ifpdf}
\usepackage{todonotes}
\usepackage{booktabs} % For formal tables
\usepackage{url}
\usepackage{pifont}% http://ctan.org/pkg/pifont
\usepackage{tcolorbox}
\usepackage{balance}
\usepackage{makecell}
\usepackage{verbatim}
\usepackage{rotating}

\usepackage{bussproofs}
\usepackage{lipsum} % just for the example
\usepackage{mathpartir}
\usepackage{algorithmic}
\usepackage{listings}
\usepackage{array}

\usepackage{multirow}
\usepackage{float}
\usepackage{mwe}
\newcommand{\code}[1]{{\fontfamily{cmtt}\fontseries{m}\fontshape{n}\selectfont\small{#1}}}

\usepackage{url}
\usepackage{listings}
\usepackage{todonotes}
\usepackage{subfigure}

\usepackage{tikz}
\usepackage{xcolor}
\definecolor{dkgreen}{rgb}{0,0.6,0}

\usepackage[para]{threeparttable}

\makeatletter
\newcommand{\linebreakand}{%
  \end{@IEEEauthorhalign}
  \hfill\mbox{}\par
  \mbox{}\hfill\begin{@IEEEauthorhalign}
}
\makeatother

\AtBeginDocument{%
  \providecommand\BibTeX{{%
    \normalfont B\kern-0.5em{\scshape i\kern-0.25em b}\kern-0.8em\TeX}}}

\renewcommand\footnotetextcopyrightpermission[1]{}

\begin{document}

\title{Characterising Bugs in Jupyter Platform}

\author{Yutian Tang}
\authornote{Corresponding author. Email: \texttt{yutian.tang@glasgow.ac.uk}}
\affiliation{%
  \institution{University of Glasgow}
  \country{United Kingdom}
}

\author{Hongchen Cao}
\affiliation{%
  \institution{ShanghaiTech University}
  \country{China}
}

\author{Yuxi Chen}
\affiliation{%
  \institution{University of Glasgow}
  \country{United Kingdom}
}

\author{David Lo}
\affiliation{%
  \institution{Singapore Management University}
  \country{Singapore}
}



% Yutian Tang
% Hongchen Cao
% Peilin He
% David Lo
% Yumin Zhou


\begin{abstract}
As a representative literate programming platform, Jupyter is widely adopted by developers, data analysts, and researchers for replication, data sharing, documentation, interactive data visualization, and more. Understanding the bugs in the Jupyter platform is essential for ensuring its correctness, security, and robustness. Previous studies focused on code reuse, restoration, and repair execution environment for Jupyter notebooks. However, the bugs in Jupyter notebooks' hosting platform Jupyter are not investigated. In this paper, we investigate 387 bugs in the Jupyter platform. These Jupyter bugs are classified into 11 root causes and 11 bug symptoms. We identify 14 major findings for developers. More importantly, our study opens new directions in building tools for detecting and fixing bugs in the Jupyter platform.

\end{abstract}

\maketitle

\section{Introduction}
Literate programming is the style of programming by interleaving executable code snippets, text descriptions, and computation results \cite{Knuth:1984}. The computation results are generated by executing the code snippets. The text description explains the source code and computation results. Such a diagram contributes to code understanding and makes the computation results explainable. 

The \textbf{Jupyter} platform is the most widely used platform for interactive literate programming \cite{shen2014interactive, Lorena:2021}. Upon the Jupyter platform, developers and data analyzers can develop \textbf{Jupyter notebooks}. They are mainly used for \textit{replication}, \textit{sharing}, \textit{documentation}, \textit{interactive data visualization} for data analysis~\cite{Brian:2021, Hans:2021, Marijan:2021}. The Jupyter platform offers the execution environment for running Jupyter notebooks, which are written by notebook authors or developers. As shown by the statistics, in September 2018, there are more than 2.5 million Jupyter notebook repositories, which are 10 times more than that in 2015 \cite{rule2018}. The Jupyter platform originated from IPython \cite{Perez:2007}. By now, more languages are supported by the Jupyter platform, such as R, JavaScript, and C.

\noindent\textbf{Motivation.} As all Jupyter notebooks are executed with the Jupyter platform, ensuring the correctness of the Jupyter platform is crucial for using and developing Jupyter notebooks. Unfortunately, the nature of bugs in the Jupyter platform is currently not well understood. It is not clear what are the root causes of the bugs in the Jupyter platform, the consequences of Jupyter bugs, and their impacts. Such information can assist Jupyter platform developers and researchers in (1) understanding the root causes of the bugs in the Jupyter platform (Finding 1-4 in Sec.\ref{subsec:rq1}); (2) understanding the common symptoms caused by bugs in the Jupyter platform (Finding 5-7 in Sec. \ref{subsec:rq2}); (3) quick localization and fixing of Jupyter bugs (Finding 11-14 in Sec.\ref{subsec:rq4} and \ref{subsec:rq5}); and (4) the development of Jupyter bug detection and testing tools (Sec. \ref{sec:suggestions}). Thus, it is time to investigate the bugs in the Jupyter platform.

\noindent\textbf{Related Work.} The existing studies explored the code replication and code reuse on Jupyter notebooks \cite{Pimentel:2019,Koenzen:2020,Wang:2020ASE, Sheeba:2021}, how to restore and repair Jupyter notebooks for reproduction \cite{Zhu:2021,Wang:2021}, code quality on Jupyter notebooks \cite{Wang:2020}. Some other non-SE domain researches care about interaction and data visualization in Jupyter notebook \cite{Li:2021,Kang:2021,koop:2017,Kery:2018,Rehman:2019,Rule:2018,Weinman:2021,Nguyen:2018}. In summary, none of the existing studies investigated the bugs on the Jupyter platform.

\noindent\textbf{Our Study.} To fill this gap, in this paper, we present the \emph{first} systematic study of the bugs on the Jupyter platform. To understand the nature of bugs in the Jupyter platform, in this paper, we aim at answering the following research questions (RQ).

\noindent$\bullet$ RQ1: What are common root causes and how often do they occur?

\noindent$\bullet$ RQ2: What are common symptoms and how often do they occur?

\noindent$\bullet$ RQ3: What are the connections between root causes and symptoms of Jupyter platform bugs?

\noindent$\bullet$ RQ4: What are the challenges in detecting Jupyter platform bugs?

\noindent$\bullet$ RQ5: What are the challenges in fixing Jupyter platform bugs?

%\noindent\textbf{Contribution.} In summary, we make the following contributions:

%\noindent$\bullet$ To the best of our knowledge, this is the \emph{first} systematic study on the Jupyter bugs;

%\noindent$\bullet$ We provide a classification of root causes and symptoms of bugs, and the connections between root causes and symptoms;

%\noindent$\bullet$ Our study offers 3 handsful suggestions for the Jupyter platform developers and calls the needs for tools for testing repairs, tools for detecting, fixing conflicts within third-party library dependencies, and tools for detecting front-end page rendering errors; and

%\noindent$\bullet$ We release the dataset in our study to allow other researchers to replicate and reproduce our study. Artifacts can be found at \cite{artifact}.

%\noindent\textbf{Skeleton.} The rest of the paper is organized as follows: Sec. \ref{sec:background} introduces the background of Jupyter system. Sec. \ref{sec:methodology} introduces the methodology we adopted for conducting this systematically research. We conduct the systematical study and answer the aforementioned research questions in Sec. \ref{sec:systematical-study}. The lessons learnt and discussions are presented in Sec. \ref{sec:discussion}. Finally, we present the related work in Sec. \ref{sec:related-work} and conclude the paper in Sec. \ref{sec:conclusion}.

\section{Background}\label{sec:background}

Knuth \cite{Knuth:1984} introduced literate programming by interleaving the code snippets and natural language to support developers in understanding the underlying thoughts behind the program segments. As a practice of literate programming, the \textit{interactive literate programming environment}, Jupyter platform, is widely adopted by developers, data analyzers, and researchers. 

Jupyter notebooks, the interactive literate programming documents, are designed and developed by developers to replicate, share, and visualize data. To avoid ambiguity, in this paper, we use the term \textit{Jupyter platform} to refer to the interactive literate programming environment. The terms \textit{Jupyter notebooks} refer to the documents developed by developers and executed upon the programming environment Jupyter platform.

\subsection{Jupyter Notebook}
A notebook is a sequence of \textit{cells} \cite{Jupyter}. The type of a cell can be a \textit{code cell} or a \textit{markdown cell}. The \textit{code cell} contains executable source code and can be used to produce results. The \textit{markdown cell} contains rich formatted texts which support the markdown. 

\begin{figure}[!htpb]
    \centering
    \includegraphics[width=0.45\textwidth]{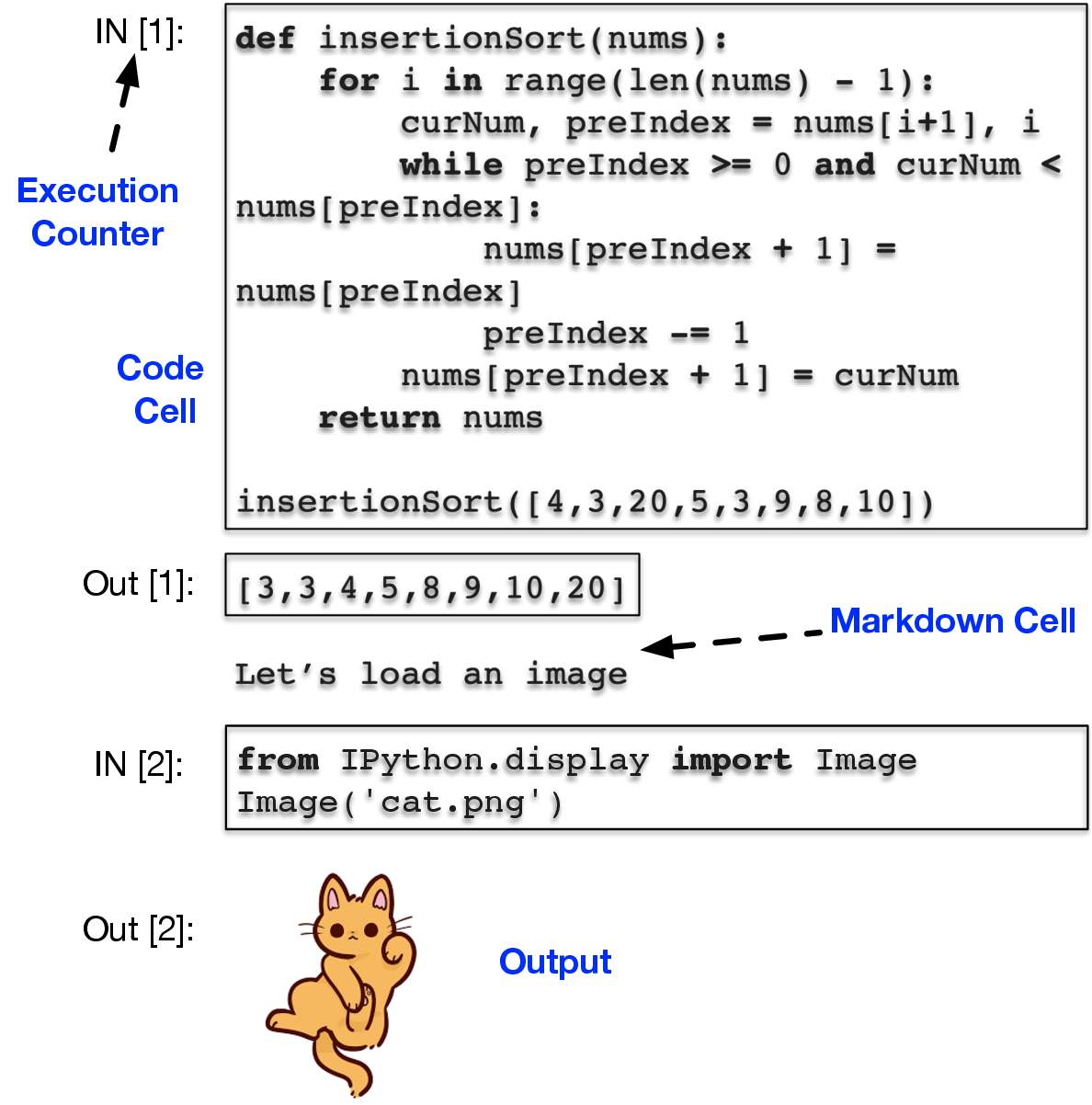}
    \vspace{-1em}
    \caption{Jupyter Notebook Example}
   \vspace{-1em}
    \label{fig:juypternotebook-example}
\end{figure}

Fig. \ref{fig:juypternotebook-example} illustrates a Jupyter notebook. It contains one markdown cell and two code cells. On the left of cells, there are \textit{execution counters}, which indicate the execution order of these cells. When these cells are executed, the outputs are displayed afterward. However, execution order only indicates the order of these cells rather than any logical relationship between cells. 

\subsection{Jupyter Platform}\label{subsec:jupyter}
\begin{figure}[!htpb]
    \centering
    \includegraphics[width=0.45\textwidth]{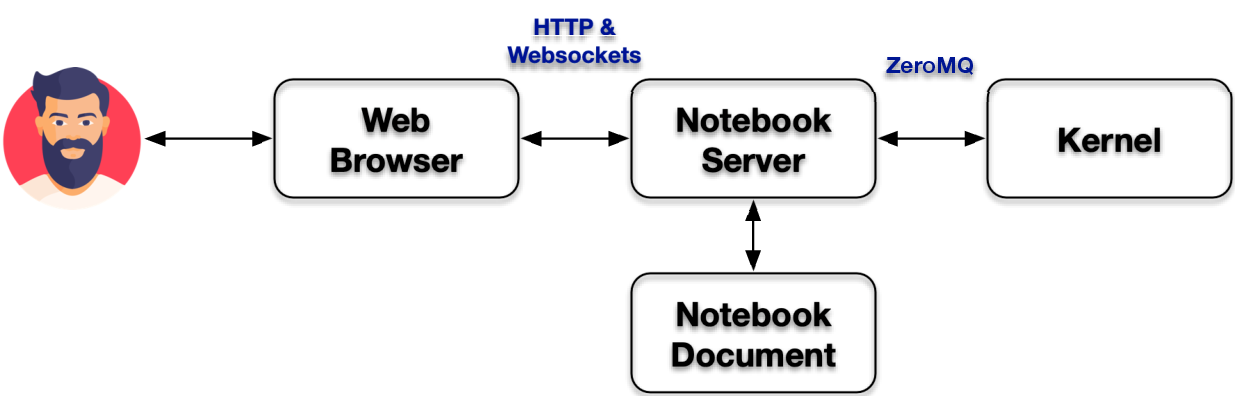}
    \vspace{-1em}
    \caption{Jupyter Architecture}
    \vspace{-1em}
    \label{fig:juypternotebook-architecture}
\end{figure}

The Jupyter platform composes of four parts: a web browser, a notebook server, a notebook document, and a kernel \cite{Jupyter}. The overall architecture is shown in Fig. \ref{fig:juypternotebook-architecture}. 

%\begin{figure}[!htpb]
%    \centering
%    \includegraphics[width=0.5\textwidth]{img/notebookdep.png}
%    \caption{Jupyter dependency graph}
%    \label{fig:juypternotebook-dependency}
%\end{figure}

\noindent\textbf{Web Browser.} Users interact with the web browser. The web browser can be considered as the frontend for the user. With this, users can manipulate notebooks.

\noindent\textbf{Notebook document.} A notebook is a file encoded with JSON, whose extension is \code{.ipynb}. The notebook can be displayed with the web browser. The code snippets, texts, and other markdown notes are stored in the editable notebook. 

\noindent\textbf{Notebook Server.} When users interact with the web browser, requests are sent to the notebook server. Requests can be either HTTP or WebSocket requests. When users require their code snippets to be executed, the notebook server sends them to the kernel over ZeroMQ sockets \cite{ZeroMQ}. The kernel executes the code snippets and returns the results to the notebook server. Then, the notebook server returns the updated HTML page to users \cite{Jupyter}.

\noindent\textbf{Kernel.} The kernel in the Jupyter platform normally refers to \code{IPython kernel}, which is in charge of running code. For other programming languages, there are other kernels, such as IRKernel for R \cite{IRkernel} and IJulia kernel for Julia \cite{IJuliaKernel}.

\subsection{Code Repositories of Jupyter Platform}\label{subsec:arch}

\begin{table}[!htpb]
\vspace{-1em}
\caption{The key Code Repositories in Jupyter}
\vspace{-1em}
\label{tab:key-modules-jupyter}
\scalebox{0.65}{
\begin{tabular}{|c|c|c|}
\hline
\textbf{Code Repo.} & \textbf{Functionality} & \textbf{Component in Fig.2} \\ \hline
Jupytercore \cite{jupytercore}  &   Core common functionality of Jupyter platform & All \\ \hline
Jupyterclient \cite{jupyterclient} &  API for managing and communicating with  kernels & Kernel, Server  \\ \hline
IPykernel \cite{ipykernel} &  IPython kernel for Jupyter platform & Kernel \\ \hline
Jupyterserver \cite{jupyterserver} &  Backend server for Jupyter notebook & Server \\ \hline
Notebook(Rep) \cite{Notebook} &  User interface for Jupyter notebook & Web browser   \\ \hline
\end{tabular}
}
\end{table}

The key repositories in the Jupyter platform are shown in Table \ref{tab:key-modules-jupyter}, including \code{Jupytercore} \cite{jupytercore}, \code{Jupyterclient} \cite{jupyterclient}, \code{IPykernel} \cite{ipykernel}, \code{Jupyterserver} \cite{jupyterserver} and \code{Notebook(Rep)} \cite{Notebook}. The first column shows the name of the repository. The second column describes the functionality of the repository. The last column shows the relations between the code repositories and components in Fig. \ref{fig:juypternotebook-architecture}. For example, \code{Jupyterserver} \cite{jupyterserver} is the repository for implementing the server. It is worth mentioning that the term ``Notebook(Rep)'' represents the code repository in the Jupyter platform for loading and rendering a Jupyter notebook (i.e., a document). Furthermore, we also leverage the \code{pydeps} \cite{pydeps}, a Python dependency visualization library, to display the dependencies of these repositories inside the Jupyter platform. Due to the size of the generated graph, we make it accessible on our artifact \cite{artifact}.

%Furthermore, \code{Jupyterlab} \cite{JupyterLab} and its server-end \code{Jupyterlabserver} \cite{Jupyterlabserver} are the next generations of the Jupyter platform. As claimed by Jupyter officials, they are gradually replacing the existing Jupyter platform. To avoid ambiguity, in this paper, we discard them. 

%As claimed by Jupyter officials, they are gradually replacing the existing Jupyter. During this period, there exists dependency relations between the existing Jupyter and its next-generation JupyterLab. However, to avoid ambiguity, in this paper, we discard them.

%Specifically, the \code{Ipykernel} is a kernel of IPython, which represents interactive Python. Note that, \code{Jupyterlab} and its server-end \code{Jupyterlabserver} are the next generations of Jupyter Notebook. As claimed by Jupyter officials, they are gradually replacing the existing Jupyter. During this period, there exists dependency relations between the existing Jupyter and its next-generation JupyterLab. However, to avoid ambiguity, in this paper, we do not consider these two components as the key components for analysis. Thus, they are excluded from our study.  

\begin{figure*}[!htpb]
    \centering
    \includegraphics[width=0.9\textwidth]{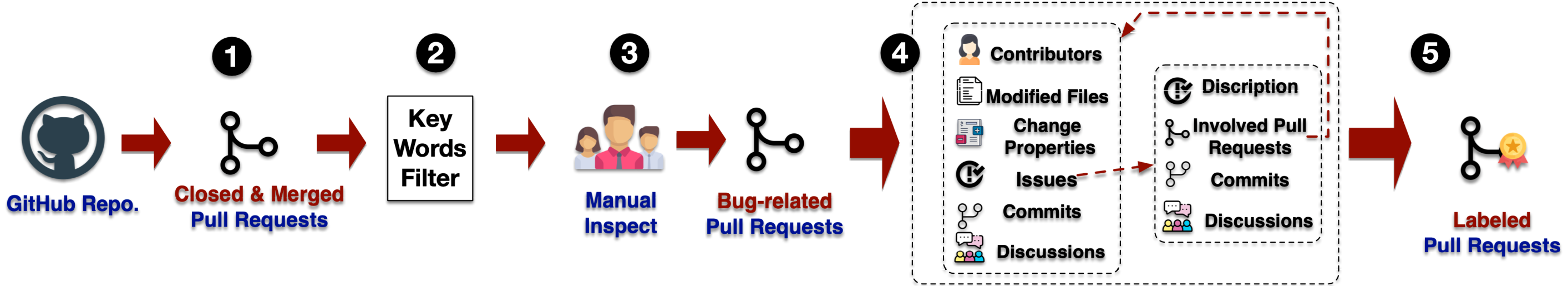}
    \vspace{-1em}
    \caption{Overview of Data Collection and Labelling}
    \vspace{-1em}
    \label{fig:juypternotebook-overview}
\end{figure*}

\section{Methodology and Classification}\label{sec:methodology}

\subsection{Data Collection, Labelling and Classification}\label{sec:classify-label}

To collect and label the data used in our study, we propose a semi-automatic approach as shown in Fig. \ref{fig:juypternotebook-overview}. Specifically, in step \ding{202}, we collect the \textit{closed} and \textit{merged} pull requests that fix bugs from the aforementioned five repositories that were created on or before March 15, 2022 via GitHub APIs. The \textit{closed} and \textit{merged} pull requests (PR) indicate that the bugs have been fixed by developers. To assist us in better understanding the fixed PRs, we also require that the selected pull requests must be associated with Issues. To filter out non bug-fixing PRs, we follow the guidance in the existing research \cite{Garcia:2020, Franco:2017,Zhang:2018,Vasilescu:2015}. Thus, in step \ding{203}, we reserve a list of words that have similar meanings to ``bugs''. The list contains: fix, defect, error, bug, issue, mistake, incorrect, fault, and flaw. Next, we search each word in both tags and titles. If a PR contains at least one keyword, we consider it as a bug-fixing PR. As a result, we obtain 510 bug-fix PRs in total. To eliminate false positives, in step \ding{204}, we manually inspect these 510 PRs and related Issues to reduce false positives. We remove some PRs that do not relate to bugs in the project. We result in 387 PRs,  with 10 PRs from Jupytercore; 30 from Jupyterclient; 56 from Jupyterserver; 57 from IPykernel; and 234 from Notebook(Rep). To characterize Jupyter bugs, we focus on labeling them from two perspectives: (1) the \textbf{root causes} that reflect the errors made by developers; (2) the \textbf{symptoms} that the bugs exhibit as represented by incorrect behaviors. Thus, in step \ding{205}, for each PR, we collect the following information: contributors, modified files, change properties (e.g., how the files are modified), Issues linked to this PR, all related commits, and discussions under this PR. Furthermore, we also collect the data for any Issue linked to this PR, including the Issue's description, all PRs related to the Issue, all related commits, and discussions under the Issue. As an Issue can map to multiple PRs and a PR can be related to multiple Issues, we iteratively crawl all related Issues and PRs. The collected data can assist us in labeling and classifying these PRs. 

\noindent\textbf{Taxonomy.} To reduce bias during classification, we ask two authors of this paper to analyze PRs separately (step \ding{206}) to label the PRs' root causes and symptoms. If there is a conflict, another author is required to label the bug. To label and classify PRs, in this paper, we reference and adjust the taxonomy used in the existing research \cite{Garcia:2020,Seaman:2008,Thung:2012} to suit the Jupyter platform. Furthermore, we adopt an open-coding scheme \cite{Garcia:2020} to expand the list of the root causes. That is, if a bug does not fall into any category, the author does a manual analysis to identify its label for the root causes. By doing this, we can expand the list of root causes. A similar procedure is conducted to set up the taxonomy of the symptoms. 

\noindent\textbf{Implementation.} We implement Python scripts to automate step \ding{202}, \ding{203}, and \ding{205}. 

\subsection{Root Causes}\label{subsec:method-cause}
According to the process described in Sec.~\ref{sec:classify-label}, the root causes are the following 11 categories:

\noindent$\bullet$ \textbf{Algorithm/Method (Alg/Meth):} The logic in the implementation of an algorithm or method is incorrect. For example, \cite{notebook-pull-799} shows an incorrect implementation of the code cell replacement function.

\noindent$\bullet$ \textbf{Assignment/Initialization (Ass/Ini):} A variable is incorrectly assigned, or mishandling of the initializations (e.g., class initialization function error, attribute assignment error in \code{.css} file). For example, the bug \cite{notebook-pull-4236} is caused by lacking of settings for \code{padding-top} in an \code{.less} file which is responsible for page rendering.

\noindent$\bullet$ \textbf{Checking:} Lack of necessary checks that lead to an error. Checking errors can be explicit (e.g., missing \code{try...catch..},\code{if...else} statements) or implicit. Taking the bug\cite{ipykernel-pull-390} as an example, as shown in List.~\ref{lst:Ipykernel390}, the developer adds \code{flush()} instead of \code{try...catch..} or \code{if...else} in line\#5 to check the status of the executed cells, which belongs to this category.

\begin{lstlisting}[label=lst:Ipykernel390, caption= An Implicit \textit{Checking} Example from Ipykernel\#390, basicstyle=\small, language=python]
def dispatch_shell(self, stream. msg):
    ...
    self._publish_status(...)
+  // flush to ensure reply is sent before handling the request
+   stream.flush(zmq.POLLOUT)
\end{lstlisting}

\noindent$\bullet$ \textbf{Logic:} Incorrect condition expressions lead to an error. For example, the bug \cite{notebook-pull-1652} is caused by using the incorrect variable (i.e., use the key of the dictionary instead of the corresponding value) in the conditional expression.

\noindent$\bullet$ \textbf{Data:} Incorrect manipulation of data items, such as incompatible types in assignments, inappropriate class inheritance, wrong type conversions, and incorrect definitions of data structures.
For example, the bug \cite{notebook-pull-2159} is caused by incorrect type conversion.

\noindent$\bullet$ \textbf{External Interface (Exter-API):} Misuse of third-party libraries or interfaces from other systems, such as incorrect function parameters passing, and invocations of deprecated functions. For example, the bug \cite{jupyterclient-pull-703} is caused by using deprecated APIs from the third-party library ZMQ~\cite{ZeroMQ}.

\noindent$\bullet$ \textbf{Internal Interface (Inter-API):} Misuse of interfaces from other components of the Jupyter platform, such as incorrect function parameters passing, and invoking inappropriate functions. For example, the bug \cite{notebook-pull-6160} is caused by passing incorrect parameters when invoking an API in Notebook(Rep).

\noindent$\bullet$ \textbf{Timing/Performance (Time/Perf): } Timing or performance problems, such as race condition, misuse of asynchronous/synchronous, and inappropriate use of multi-threading. For example, the bug \cite{jupyterclient-pull-607} is caused by a race condition between the restart module and the shutdown module in the Jupyterclient.

\noindent$\bullet$ \textbf{Configuration (Config):} Misconfiguration of files for compilation, build, test, and installation (e.g., incompatible third-party library versions, and inappropriate package importation). For example, the bug \cite{jupyterserver-pull-521} is caused by using an outdated third-party library anyio~\cite{anyio}, which is incompatible with Python 3.6.

\noindent$\bullet$ \textbf{Non-functional (Non):} Non-functional errors that do not directly affect the use of the Jupyter platform (e.g., inappropriate description in error traceback and log). For example, the bug \cite{notebook-pull-5136} is caused by incorrect log messages which can make users confused.

\noindent$\bullet$ \textbf{Others:} Other root causes that cannot be classified into any of the above categories. For example, the bug \cite{ipykernel-pull-18} is caused by \code{Datetime} objects do not follow time data standard ISO8601~\cite{iso8601}.

\subsection{Symptoms}\label{sec:symptom-label}
The complete list of symptoms is in the following 11 categories:

\noindent$\bullet$ \textbf{Crash: } Critical errors that lead to a crash at runtime. For example, Issue \cite{notebook-issue-96} reports that the web browser crashes after the user uploads large files to the Jupyter platform.

\noindent$\bullet$ \textbf{Hang: } The kernel (i.e., IPykernel) or terminal (e.g., terminal in Jupyterclient) is not responding (e.g., fails to kill the kernel, infinite loop). For example, Issue \cite{ipykernel-issue-742} reports that running certain code (i.e., \code{\%gui tk}) makes the IPykernel has no response and gets stuck.

\noindent$\bullet$ \textbf{Build: } The errors during installation or building. For example, the installation failure reported in \cite{ipykernel-issue-694} is due to the lack of the dependency library \code{ipyparallel}~\cite{ipyparallel} in IPykernel.

\noindent$\bullet$ \textbf{Display and GUI (DGUI): } The GUI-related or display-related errors (e.g., missing widgets, misalignment of the icons and fonts). For example, a long line of code can be overlapped with the border of the code cell~\cite{notebook-issue-1003}.

\noindent$\bullet$ \textbf{Launch: } Any errors that occur during Launch. For example, the user fails to open a notebook (i.e., \code{.ipynb} file) after setting the environment variable (i.e., \code{JUPYTER\_PATH}) to a custom value~\cite{jupyterclient-issue-591}.

\noindent$\bullet$ \textbf{IO: }
Incorrect behaviors when performing inputs/outputs to and interacting with the Jupyter platform. For example, the user cannot delete empty lines by pressing Backspace key on the keyboard \cite{notebook-issue-5502}.

\noindent$\bullet$ \textbf{Security and Safety (SS): } Errors cause security vulnerabilities that can be exploited, such as incorrect permissions and information leakage.
For example, the exploitable vulnerability reported in ~\cite{notebook-pull-22} can lead to the leakage of the user information.

\noindent$\bullet$ \textbf{Test: } Errors found during the testing phase by causing test suites to fail, usually have no direct symptom description. 

\noindent$\bullet$ \textbf{Unreported (Un): } Error in which bug reporters do not give a clear symptom description. For example, the user only reports that the Jupyter platform does not work with tornado6~\cite{tornado} but does not give any further description~\cite{jupyterserver-issue-42}.

\noindent$\bullet$ \textbf{Deprecation (De): } The errors related to outdated third-party libraries/modules. For example, PR \cite{jupyterserver-issue-591} shows that using a deprecated regex API~\cite{re} triggers a warning message from Jupyterserver.

\noindent$\bullet$ \textbf{Others: } Other symptoms that cannot be regarded as one of the above categories. For example, the notebook (i.e., \code{.ipynb} file) is cleared after renaming the file~\cite{notebook-issue-5190}.

\subsection{Research Questions (RQ)}
Our study aims to answer the following five research questions:

\noindent$\bullet$ \textbf{RQ1: What are common root causes and how often do they occur?} Studying root causes contributes to the understanding of the nature of bugs in the Jupyter platform. The classification of root causes and the statistics of their distribution are the basis for further analysis. 

\noindent$\bullet$ \textbf{RQ2: What are common symptoms and how often do they occur?} Symptoms are the intuitive expression of errors, which provide developers with a hint for solving the errors. Exploring symptoms helps developers understand the Jupyter platform's bugs. 

\noindent$\bullet$ \textbf{RQ3: What are the connections between root causes and symptoms of Jupyter platform bugs?} Existing studies show that the root causes and the symptoms of bugs are intrinsically related~\cite{Garcia:2020, Islam:2019, Thomas:2003}. Learning whether a  root cause can be linked to a symptom can help developers to further understand and detect Jupyter bugs. 

\noindent$\bullet$ \textbf{RQ4: What are the challenges in detecting Jupyter platform bugs?} Detecting bugs is an essential prerequisite for bug analysis and fixing. In this RQ, we analyze the challenges in bug detection.

\noindent$\bullet$ \textbf{RQ5: What are the challenges in fixing Jupyter platform bugs?} After successfully locating the source of the bug, how to fix it efficiently is the final hurdle. It allows developers to fix bugs more efficiently.

%hongchen: RQ1 RQ5 RQ3
%peilin: RQ2 RQ4  RQ3

\section{Results}\label{sec:systematical-study}
\subsection{RQ1: Root Causes}\label{subsec:rq1} % Root Tag data

\noindent\textbf{Methodology.}
The methodology is described in Sec.~\ref{sec:methodology}.

\begin{table}[!htbp]
%\vspace{-1em}
\caption{Distribution of Root Causes }
%\vspace{-1em}
\label{tab:root-cause}
\scalebox{0.8}{
\begin{threeparttable}
\begin{tabular}{lcccccc}
\hline
\textbf{Root Cause}                 & \textbf{J.core}\tnote{1} & \textbf{J.client}\tnote{1} & \textbf{IPykernel} & \textbf{J.server}\tnote{1} & \textbf{Note.}\tnote{1} & \multicolumn{1}{l}{$Total_{Cause}$} \\ \hline
Alg/Meth         & 1                    & 6                      & \textcolor{blue}{\textbf{20}}                 & \textcolor{blue}{\textbf{15}}                     & 54                & \textcolor{red}{\textbf{96(24.81\%)}}                                            \\ \hline
Ass/Ini & 2                    &    4         & 9                  & 7                      & \textcolor{blue}{\textbf{67}}                & \textcolor{red}{\textbf{89(23.00\%)}}                                            \\ \hline
Checking                            & \textcolor{blue}{\textbf{4}}                   & \textcolor{blue}{\textbf{7}}                      & 11                 & 9                      & 32                & \textcolor{red}{\textbf{63(16.28\%)}}                                            \\ \hline
Logic                               & 0                    & 5                      & 4                  & 3                      & 23                & \textbf{35(9.04\%)}                                            \\ \hline
Data                                & 0                    & 0                      & 2                  & 0                      & 1                 & \textbf{3(0.78\%)}                                             \\ \hline
Exter-API      & 1                    & 2                      & 3                  & 2                      & 2                 & \textbf{10 (2.58\%)}                                             \\ \hline
Inter-API     & 0                    & 1                      & 1                  & 1                      & 7                 & \textbf{10 (2.58\%)}                                            \\ \hline
Time/Perf      & 0                    & 1                      & 0                  & 1                      & 2                 & \textbf{4 (1.03\%)}                                             \\ \hline
Config              & 2                    & 2                      & 5                  & \textcolor{blue}{\textbf{15}}                     & 37                & \textcolor{red}{\textbf{61 (15.76\%)}}                                            \\ \hline
Non                & 0                    & 0                      & 1                  & 2                      & 6                 & \textbf{9 (2.32\%)}                                             \\ \hline
Others                              & 0                    & 2                      & 1                  & 1                      & 3                 & \textbf{7 (1.81\%)}                                            \\ \hline
\textit{$Total_{Module}$}          & \textbf{10}          & \textbf{30}            & \textbf{57}        & \textbf{56}            & \textbf{234}      & \textbf{387}                                \\ \hline 
\end{tabular}
\end{threeparttable}
}
\begin{tablenotes}
    \small \item [1] J.core for Jupytercore, J.clicent for Jupyterclient, J.server for Jupyterserver, \\ Note. for Notebook(Rep)

\end{tablenotes}
\end{table}

\noindent\textbf{Results.} Table.~\ref{tab:root-cause} shows the distribution of Jupyter platform bugs by root cause categories. For each component of the Jupyter platform, its most frequent root cause is as follows:

\noindent$\bullet$ For Jupytercore and Jupyterclient, \textbf{Checking} is the top root cause, accounting for 40.00\% of all bugs in Jupytercore and 23.33\% of all bugs in Jupyterclient; 

\noindent$\bullet$ For IPykernel, \textbf{Alg/Meth} is the top root cause (35.08\% of all bugs in IPykernel).

\noindent$\bullet$ For Jupyterserver, \textbf{Alg/Meth} and \textbf{Config} are tied for first place (both 26.79\% of all bugs in Jupyterserver).

\noindent$\bullet$ For Notebook(Rep), \textbf{Ass/Ini} is the top root cause (28.63\% of all bugs in Notebook(Rep)).

\noindent$\bullet$ Overall, Alg/Meth, Ass/Ini, Checking, and Config are the top four frequent root causes for the Jupyter, accounting for 24.81\%, 23.00\%, 16.28\%, and 15.76\% of all studied bugs, respectively.

Alg/Meth is the most frequently occurring category. We find that for different components, the algorithms with bugs have different characteristics. Specifically, according to Fig.~\ref{fig:juypternotebook-architecture} and Table.~\ref{tab:key-modules-jupyter}, every component has its main functionality. Algorithms with bugs often have a strong relationship with the functionality of their components. Recall that Alg/Meth is the top root cause of bugs in Jupyterserver and IPykernel. For Jupyterserver, 8 out of 15 PRs modify the file \code{handlers.py}, suggesting that algorithms related to server handlers are prone to bugs. For IPykernel, half of the PRs modify \code{kernelbase.py} or \code{iostream.py}, suggesting that algorithms related to kernel read and write are not robust. 

\begin{tcolorbox}[title=Finding 1,boxrule=1pt,boxsep=1pt,left=2pt,right=2pt,top=2pt,bottom=2pt]
Alg/Meth is the most frequent root cause, accounting for 24.81\% of Jupyter platform bugs. 
\end{tcolorbox} 

Ass/Ini is the second most frequent root cause. As defined in Sec.~\ref{subsec:method-cause}, this category involves the assignment or initialization errors or mishandling of the initializations. Recall that Ass/Ini is the top root cause of bugs in Notebook (Rep).
Notebook(Rep) involves front-end browser pages that interact with users and consist of files in the format of \code{.html}, \code{.css}, \code{.js}, \code{.less}. It can be error-prone to manually set up all attributes for front-end UI elements in Notebook(Rep). Even simple mistakes (i.e., Ass/Ini bugs) can lead to various errors, such as page rendering errors, front-end interaction failures, and other problems. In addition, some page rendering errors are difficult to find through tests and are often noticed by users or other developers. Taking Notebook(Rep)'s PR\#4236~\cite{notebook-pull-4236} as an example, an icon is overlaid on the docstring of the tooltip. Such an incorrect layout prevents the user from reading the full description of the function signature. The difference between the normal and incorrect UI is so subtle that it can be easily overlooked.

\begin{tcolorbox}[title=Finding 2,boxrule=1pt,boxsep=1pt,left=2pt,right=2pt,top=2pt,bottom=2pt]
Ass/Ini is the second most frequent root cause, accounting for 23.00\% of Jupyter platform bugs. This category of bugs is rampant in Notebook(Rep) and sometimes hard to detect.
\end{tcolorbox} 

Checking is the third most frequent root cause. This category of bugs is relatively trivial to fix, usually by adding \code{if...else} or \code{try...catch}. 
Recall that Checking is the top root cause of bugs in Jupytercore or Jupyterclient.

\underline{Jupytercore} offers the core common functionality of the Jupyter platform. Thus, it involves a significant amount of checks, such as checks for inputs, configurations, and options. Lacking such checks can make the Juypter platform fail to work properly. For example, PR \cite{jupytercore-pull-183} reports a bug caused by lacking file permission checks on Windows. \underline{Jupyterclient} offers APIs for starting, managing, and communicating with Jupyter kernels. When invoking these APIs, developers have to pass arguments to these APIs. Inside these APIs, sufficient checks must be conducted on these passed arguments. Insufficient checks can lead to exceptions. For example, PR \cite{jupyterclient-pull-254} reports a bug due to the lack of checking whether the \code{meth.\_\_doc\_\_} is \code{None} or not.

\begin{tcolorbox}[title=Finding 3,boxrule=1pt,boxsep=1pt,left=2pt,right=2pt,top=2pt,bottom=2pt]
Checking is the third most frequent root cause, accounting for 16.28\% of Jupyter platform bugs. Under this category, exception-related bugs are potentially detectable and fixable by existing tools.
\end{tcolorbox} 

Config is also a serious root cause that occupies a significant percentage. As defined in Sec.~\ref{subsec:method-cause}, this category involves misconfiguration of files, such as files for compilation and build. Recall that Config is the top root cause of bugs in Jupyterserver. According to the dependency graph of the Jupyter platform (see our artifact~\cite{artifact}), Jupyterserver has 14 upstream dependencies. The most common Config errors in Jupyterserver are the incorrect settings of third-party libraries' versions, such as traitlets~\cite{Traitlets}, anyio~\cite{anyio}. As soon as the third-party dependency library conflicts occur, the developer needs to change the configuration of the related libraries. Modifying the configuration is not as thorny as fixing code-related issues. However, it is hard to quickly locate the library causing the conflict among the vast number of dependency libraries. 

\begin{tcolorbox}[title=Finding 4,boxrule=1pt,boxsep=1pt,left=2pt,right=2pt,top=2pt,bottom=2pt]
Config is another serious root cause, accounting for 15.76\% of Jupyter platform bugs. The incorrect settings of third-party libraries' versions are the source of Config bugs.
\end{tcolorbox} 

\subsection{RQ2: Symptoms}\label{subsec:rq2} % Symptoms Tag data

\noindent\textbf{Methodology.} The detailed methodology is described in Sec.~\ref{sec:methodology}.

\begin{table}[!htbp]
\vspace{-1em}
\caption{Distribution of Symptoms }
\vspace{-1em}
\label{tab:Symptom}
\scalebox{0.8}{
\begin{threeparttable}
\begin{tabular}{lcccccc}
\hline
\textbf{Symptom}                 & \textbf{J.core} & \textbf{J.client} & \textbf{IPykernel} & \textbf{J.server} & \textbf{Note.} & \multicolumn{1}{l}{$Total_{Symptom}$} \\ \hline
Build         & 2                    & 1                      & 4                 & 2                     & 13                & \textbf{22 (5.68\%)}                                            \\ \hline
Crash & 1                    & 4                      & 2                  & 2                      & 4                & \textbf{13 (3.36\%)}                                            \\ \hline
DGUI                            & 1                    & 0                      & 4                 & 2                      & 66                & \textcolor{red}{\textbf{73 (18.86\%)}}                                            \\ \hline
Hang                               & 0                    & 3                      & 6                  & 6                      & 2                & \textbf{17 (4.39\%)}                                            \\ \hline
IO                                & 2                    & 6                      & \textcolor{blue}{\textbf{21}}                  & 8                      & \textcolor{blue}{\textbf{78}}                 & \textcolor{red}{\textbf{115 (29.72\%)}}                                             \\ \hline
Launch      & \textcolor{blue}{\textbf{3}}                    & \textcolor{blue}{\textbf{8}}                      & 3                  & \textcolor{blue}{\textbf{14}}                  & 39                 & \textcolor{red}{\textbf{67 (17.31\%)}}                                             \\ \hline
SS     & 0                    & 0                      & 0                  & 0                      & 4                 & \textbf{4 (1.03\%)}                                            \\ \hline
Test      & 1                    & 2                      & 6                  & 7                      & 11                 & \textbf{27 (6.98\%)}                                             \\ \hline
Un              & 0                    & 2                      & 2                  & 8                     & 8                & \textbf{20 (5.17\%)}                                            \\ \hline
De                              & 0                    & 2                      & 3                  & 2                      & 2                 & \textbf{9 (2.33\%)}                                            \\ \hline
Others                              & 0                    & 2                      & 6                  & 5                      & 7                 & \textbf{20 (5.17\%)}                                            \\ \hline
\textit{$Total_{Module}$}          & \textbf{10}          & \textbf{30}            & \textbf{57}        & \textbf{56}            & \textbf{234}      & \textbf{387}                                \\ \hline 
\end{tabular}
\end{threeparttable}
}
\begin{tablenotes}
    \small \item [1] J.core for Jupytercore, J.clicent for Jupyterclient, J.server for Jupyterserver, \\ Note. for Notebook(Rep)
\end{tablenotes}
\end{table}

\begin{table*}[!htbp]
%\vspace{-1em}
\caption{Connections between Root Causes and Symptoms}
%\vspace{-1em}
\label{tab:relation}
\scalebox{0.8}{
\begin{tabular}{lcccccccccccc}
\hline
\multicolumn{1}{c}{} & \textbf{Crash}               & \textbf{Hang}                & \textbf{Build}               & \textbf{Launch}              & \textbf{IO}                  & \textbf{DGUI}             & \textbf{SS}             & \textbf{Test}                & \textbf{Un}                & \textbf{De}                & \textbf{Others}               & $Total_{Causes}$             \\ \hline
\textbf{Alg/Meth}    & 4          & 10          & 1          & 14          & \textcolor{blue}{\textbf{40 (41.67\%)}}          & 13          & 1          & 4          & 4          & 0          & 5          & \textbf{96} \\ \hline
\textbf{Ass/Ini}     & 1          & 1          & 3          & 13          & 24          & \textcolor{blue}{\textbf{41 (46.07\%)}}          & 0          & 2          & 2          & 0          & 2          & \textbf{89} \\ \hline
\textbf{Checking}    & 5          & 4          & 4          & 15          & \textcolor{blue}{\textbf{19 (30.16\%)}}          & 3          & 3          & 5          & 1          & 0          & 4          & \textbf{63} \\ \hline
\textbf{Data}        & 0          & 0          & 0          & 0          & 0          & 2          & 0          & 0          & 1          & 0          & 0          & \textbf{3} \\ \hline
\textbf{Exter-API}   & 1          & 0          & 0          & 0          & 0          & 0          & 0          & 0          & 0          & 9          & 0          & \textbf{10} \\ \hline
\textbf{Inter-API}   & 1          & 0          & 0          & 2          & 5          & 1          & 0          & 0          & 1          & 0          & 0          & \textbf{10} \\ \hline
\textbf{Logic}       & 1          & 0          & 1          & 9          & 15          & 3          & 0          & 2          & 2          & 0          & 2          & \textbf{35} \\ \hline
\textbf{Non}         & 0          & 0          & 1          & 1          & 0          & 2          & 0          & 0          & 1          & 0          & 4          & \textbf{9} \\ \hline
\textbf{Time/Perf}    & 0          & 0          & 0          & 1          & 2          & 0          & 0          & 1          & 0          & 0          & 0          & \textbf{4} \\ \hline
\textbf{Config}      & 0          & 2          & \textcolor{blue}{\textbf{12 (19.67\%)}}          & 11          & 9          & 8          & 0          & 10          & 7          & 0          & 2          & \textbf{61} \\ \hline
\textbf{Others}      & 0          & 0          & 0          & 1          & 1          & 0          & 0          & 3          & 1          & 0          & 1          & \textbf{7} \\ \hline
$Total_{Symptoms}$   & \textbf{13} & \textbf{17} & \textbf{22} & \textbf{67} & \textbf{115} & \textbf{73} & \textbf{4} & \textbf{27} & \textbf{20} & \textbf{9} & \textbf{20} & \textbf{387} \\ \hline
\end{tabular}
}
\end{table*}

\noindent\textbf{Results.}
Table.~\ref{tab:Symptom} shows the distribution of different symptoms for Jupyter platform bugs.

\noindent$\bullet$ For Jupytercore, Jupyterclient, and Jupyterserver, \textbf{Launch} is the top frequent symptom, accounting for 30.00\% of all bugs in Jupytercore, 26.67\% of all bugs in Jupyterclient, and 25.00\% of all bugs in Jupyterserver. 

\noindent$\bullet$ For IPykernel, Notebook(Rep), \textbf{IO} is the top frequent symptom, accounting for 36.84\% of all bugs in IPykernel and 33.33\% of all bugs in Notebook(Rep).

\noindent$\bullet$ Furthermore, IO, DGUI, and Launch are the top three frequent symptoms, accounting for 29.72\%, 18.86\%, and 17.31\% of all studied bugs, respectively.

Among all symptoms, IO is the most frequently reported symptom. Specifically, the IO symptom appears in all components. Recall that IO is the top symptom of bugs in Notebook(Rep) and IPykernel.

For \underline{Notebook(Rep)}, it implements the frontend for the end-users. As it directly interacts with users, the IO errors are mostly reported in Notebook(Rep). For example, PR \cite{notebook-issue-2203} reports a bug that the Edit/View buttons are either not working or working incorrectly. 

Sometimes, IOs can be complex. For example, users need to query some information from a server and run a code snippet. Such complex IO requests can only be processed with backend components (e.g.,Jupterclient, IPykernel). An error in a backend component can lead to an incorrect result, which is finally presented to end-users. That is the reason why IO errors can also be reported in backend components. For \underline{IPykernel}, as shown in Fig. \ref{fig:juypternotebook-architecture}, the users' requests are sent by Notebook Server via ZeroMQ. The \code{iostream.py} in the IPykernel is in charge of parsing the ZeroMQ messages. Error in the IPykernel (i.e., \code{iostream.py}) can return the incorrect results to users, which results in the IO symptom.

\begin{tcolorbox}[title=Finding 5,boxrule=1pt,boxsep=1pt,left=2pt,right=2pt,top=2pt,bottom=2pt]
IO is the most frequently reported symptom, accounting for 115 out of 387. The IO symptom appears in all components of the Jupyter. IO is the top frequent symptom in Notebook(Rep) and IPykernel.
\end{tcolorbox}

DGUI is the second most frequent symptom. Most DGUI errors appear in Notebook(Rep) as it is mainly in charge of interaction with end-users. Many front-end errors and rendering errors (i.e., DGUIs) are rooted in Notebook(Rep). Some DGUI problems come from the back-end components (e.g., Jupytercore, IPykernel, Jupyterserver). The reason for this is that these components are in charge of processing the requests (e.g., compiling a code segment) from the front end. Then, the final results are returned to the front end for rendering. The DGUI symptom can appear when there are some wrong implementations in these components. For example, when a user wants to query files, this request is processed by \code{FileManager} in the Juptercore and returned to the user via the web browser. Incorrect implementation of \code{FileManager} in the Juptercore can lead to an unexpected display for the user (i.e., a DGUI symptom).

\begin{tcolorbox}[title=Finding 6,boxrule=1pt,boxsep=1pt,left=2pt,right=2pt,top=2pt,bottom=2pt]
DGUI is the second most frequent symptom, accounting for 18.86\% of all bugs. The DGUI symptom appears in Jupytercore, IPykernel, Jupyterserver, and Notebook(Rep), with the most occurrence in Notebook(Rep).
\end{tcolorbox} 

%第三个经常发生的症状是Launch。Launch症状发生在Jupyter的所有component中，因为启动Jupyter平台需要启动Jupyter所有的component。任何一个component的启动错误都会导致Jupyter平台出现Launch症状。同时，我们可以注意到，在Jupytercore，Jupyterclient和Jupyterserver中，Launch症状是最常出现的症状。

The third most prominent symptom is Launch. The Launch symptom appears in all components of Jupyter. The reason for this is that the execution of the Jupyter platform involves launching all the components of Jupyter. Besides, it is worth noting that Launch is the top frequent symptom in Jupytercore, Jupyterclient, and Jupyterserver.

\noindent$\bullet $ The Launch errors are reported in Jupytercore mainly due to the lacks of checks in command-line options. In general, users can leverage the \code{jupyter notebook} command to launch the Jupyter platform. This command (i.e., \code{jupyter notebook}) offers some options (e.g.,--port, --no-browser) to support specific needs \cite{first-way-to-run-Jupyter}. Such options are processed in Jupytercore. The lacks of checks in invalid command options can lead to a Launch error when the user executes the Jupyter with such options.

\noindent$\bullet $ In Jupyterclient, Launch is the top frequent symptom. According to Table. \ref{tab:key-modules-jupyter}, Jupyterclient is in charge of managing and communicating with IPykernel. Thus, the errors in the Jupyterclient can affect the usage and launch of IPykernel (e.g., IPykernel cannot start at launch). Launching the Jupyter platform requires the IPykernel to be launched as well. The launching problem in IPykernel causes the Launch symptom. For example, PR \cite{jupyterclient-pull-717} reports a bug that the errors in the Jupyterclient make that the Jupyterclient sends multiple requests to restart the IPykernel. This is because the Jupyterclient fails to check the status (e.g., starting, closed) of the IPykernel. As a result, the Jupyter platform fails to launch. 

\noindent$\bullet $ In Jupyterserver, the most frequent symptom is also Launch. In general, we find two possible reasons:

(1) The Jupyterserver offers the backend server for the Jupyter platform. In the Jupyterserver, the ServerApp (i.e.,\code{serverapp.py} in Jupyterserver) is the core part, which connects all components in the Jupyterserver \cite{jupyterserver-architecture}. When starting the Jupyter platform, the Jupyterserver is launched as well \cite{jupyterserver-auto-launch}. The errors (e.g., incorrect assignment, failure to process command-line options) in the ServerApp (the \code{serverapp.py}) can halt the launch process of the Jupyterserver. As a result, users encounter a Launch error. The examples can be found in PR \cite{jupyterserver-pull-473} and PR \cite{jupyterserver-pull-380}.

(2) When users open a Jupyter notebook with the command \code{jupyter notebook <target>.ipynb}, the frontend of the Jupyter platform sends a POST request to the Jupyterserver \cite{jupyterserver-post-information,jupyterserver-workflow}. The detailed introduction of the POST request can be found in \cite{jupyterserver-post-information}. When receiving this request, the Mapping Kernel Manager (MKM) inside the Jupyterserver calls the Kernel Manager (KM) inside the Jupyterclient to launch the IPykernel \cite{jupyterserver-workflow} to parse the input notebook file. The errors in the MKM can make the IPykernel fails to connect and communicate with the Jupyterserver (e.g., IPykernel cannot be launched, IPykernel stops responding). As a result, an error (e.g., time out error) is reported during the launching process of the Jupyter platform. An example can be found in \cite{jupyterserver-pull-482}.

\begin{tcolorbox}[title=Finding 7,boxrule=1pt,boxsep=1pt,left=2pt,right=2pt,top=2pt,bottom=2pt]
Launch is the third most prominent symptom, accounting for 67 out of 387 symptoms.
The Launch symptom appears in every component. In Jupytercore, Jupyterclient, and Jupyterserver, Launch is the top frequent symptom that occurs most frequently.
\end{tcolorbox}

\subsection{RQ3: Connections between Root Causes and Symptoms}\label{subsec:rq3} % Root&Symptoms Tag data

\noindent\textbf{Methodology.}
Based on the statistics in RQ1 and RQ2, we further calculate relationship between root causes and symptoms. Furthermore, we combine the four most significant root causes with their highest number of symptoms into four groups. We only regard these four groups as noteworthy connections.

\noindent\textbf{Results.}
Table~\ref{tab:relation} shows the frequency relationship between root causes and symptoms. The top four noteworthy connections are:

\noindent $(1)$ The connection between Alg/Meth and IO.

\noindent $(2)$ The connection between Ass/Ini and DGUI.

\noindent $(3)$ The connection between Checking and IO.

\noindent $(4)$ The connection between Config and Build.

%根据（1）和（3），IO 相关的bug通常是在web browser被注意到。这是因为用户通过web browser和Jupyter进行交互。尽管Alg/Meth和Checking都可以导致IO症状，但是他们的表现不同。

For (1) and (3), IO bugs are observed in the frontend (i.e.,web browser), as the user interacts with Jupyter through this component.

% IO-Alg/Meth @hepl
%对于IO和Alg/Meth之间的关系，我们发现，许多问题IO的问题与Alg/Meth相关主要是因为Jupyter负责与处理用户IO请求的一系列函数中出现了错误。使得Jupyter不能正确地处理用户在web browser中的发起的大量的IO请求。更糟的是，这些错误的解决不能简单地依靠添加一些检查条件或者修改一些逻辑来解决。开发者需要完整地编写一个完整的函数，或者一系列处理算法，才能解决这些异常情况。因此，IO症状和Alg/Meth具有关联性。

For (1), Alg/Meth can be the root cause of the IO symptom. When users interact with the Jupyter platform, algorithms, and methods in the Jupyter platform are in charge of processing users' requests and returning results to users. Thus, errors in Alg/Meth can make users fail to get the correct results or even crash when interacting with the Jupyter platform. For example, PR \cite{notebook-issue-792} reports a bug in copying and pasting multiple cells. Such a bug is due to the incorrect implementation of the corresponding algorithm. Thus, Alg/Meth can be the root cause of the IO symptom.

% IO-Checking @hongchen
% 我们发现IO bug的出现往与用户的操作和环境配置有关。
For (3), Checking can also be the root cause of IO symptoms. When interacting with users, the Jupyter platform needs to perform a series of checks to ensure correctness, such as whether users' inputs are valid and acceptable. Lacking such checks can lead to unexpected outputs or errors (i.e., the IO symptom). Taking Notebook PR\#1011~\cite{notebook-pull-1011} as an example, the user discovers a bug when using Jupyter in the Firefox browser. The user reports that two same consecutive pop-ups ask users to confirm whether they want to exit after clicking the "Close and Halt" button. The reason is that the browser fires the \code{beforeunload} event in Javascript after clicking the button. Once the event is detected, a pop-up window appears. Firefox triggers this event twice when a browser window with unsaved content is about to close. While other browsers only fire the event once. Jupyter platform fails to check the user's browser type, which causes the bug. Thus, Checking can be the root cause of the IO symptom.

\begin{tcolorbox}[title=Finding 8,boxrule=1pt,boxsep=1pt,left=2pt,right=2pt,top=2pt,bottom=2pt]
For IO sympotoms, Alg/Meth and Checking can be the main root causes.
\end{tcolorbox} 

\begin{lstlisting}[label=lst:Notebook4236, caption= An \textit{Ass/Ini} Example from Notebook\#4236]
.tooltiptext{
   padding-right:30px
+ /*avoid the ui-icon(s) from overlapping the tooltip*/
+  padding-top:30px;}
\end{lstlisting}

For(2), functional modules responsible for the display and GUI rely on front-end programming languages, which involve a huge volume of settings for the properties and attributes in UI elements. Such settings are implemented with a series of assignment statements and initializations (Ass/Ini) in files (e.g., \code{.html}, \code{.css}, \code{.js}). Thus, the errors in Ass/Ini can result in page rendering errors. Taking Notebook(Rep)’s PR\#4236~\cite{notebook-pull-4236} in Sec.~\ref{subsec:rq1} as an example, the commit for fixing this bug is shown in List.~\ref{lst:Notebook4236}. The source of the bug is that the attribute \code{padding-top} is not set. As a result, such errors can lead to the DGUI symptom. In summary, for the DGUI, Ass/Ini can be the root cause.

\begin{tcolorbox}[title=Finding 9,boxrule=1pt,boxsep=1pt,left=2pt,right=2pt,top=2pt,bottom=2pt]
The bugs whose symptom is DGUI can be closely related to Ass/Ini. The massive and easily overlooked attribute assignments in front-end programming contribute to the strong connection.
\end{tcolorbox} 

For (4), the Config refers to the misconfiguration of files. The functions responsible for the installation and building rely on many configuration files. Wrong settings in such configuration files can cause Jupyter to behave incorrectly during the installation or building process, resulting in the Build symptom. For example, in Notebook(Rep) Issue\#1977 \cite{notebook-issue-1977}, a user fails to deploy the Dev. version of the Jupyter. By manually inspecting the Issue in the Notebook(Rep), we find that the incorrect configuration files lead to this error. Also, in Notebook(Rep) Issue\#52 \cite{notebook-issue-52}, incorrect configuration of Dockerfile leads to the failure of building Notebook(Rep) images. Thus, there is a relation between the Config and Build.

\begin{tcolorbox}[title=Finding 10,boxrule=1pt,boxsep=1pt,left=2pt,right=2pt,top=2pt,bottom=2pt]
Most bugs with symptoms reported as Build are related to improper configuration. Incorrect settings of the configuration files lead to the errors during the Build. It suggests a relation between the Config and Build.
\end{tcolorbox}

\begin{table}[!htbp]
%\vspace{-1em}
\caption{Ranking of Symptoms for Bug Detection}
%\vspace{-1em}
\label{tab:challenge-symptom}
\scalebox{0.75}{
\begin{threeparttable}
    \begin{tabular}{lccccc}
\hline
\multicolumn{1}{c|}{\multirow{2}{*}{\textbf{Symptom}}} & \multicolumn{5}{c}{\textbf{Issues}}                                    \\ \cline{2-6} 
\multicolumn{1}{c|}{}                                     & \#Comment & \#Participant      & Contributor(\%)          & External(\%)    & Ranking    \\ \hline
Crash                                                 & 6.92           & 3.85          & 38\%          & 31\%          & 3          \\ \hline
Hang                                                   & 4.18           & 2.47          & 59\%          & 29\%           & 7          \\ \hline
Build                                                  & 5.91           & 2.95          & 50\%          & 0\%           & 8          \\ \hline
Launch                                                 & \textcolor{blue}{\textbf{10.24}}           & \textcolor{blue}{\textbf{5.01}}          & \textcolor{blue}{\textbf{39\%}}          & \textcolor{blue}{\textbf{15\%}}           & \textcolor{blue}{\textbf{2}}          \\ \hline
IO                                                     & 5.93           & 3.34          & 55\%          & 10\%          & 6          \\ \hline
DGUI                                                & 6.37           & 3.33          & 41\%           & 14\%          & 4         \\ \hline
SS                                                 & \textcolor{blue}{\textbf{8.25}}           & \textcolor{blue}{\textbf{4.75}}          & \textcolor{blue}{\textbf{25\%}}          & \textcolor{blue}{\textbf{25\%}}          & \textcolor{blue}{\textbf{1}}          \\ \hline
Test                                                 & 4.96           & 2.26          & 85\%          & 7\%            & 9          \\ \hline
Un                                                    & 2.30        & 1.95          & 80\%          & 0\%   & 11          \\ \hline
De                                                       & 6.67           & 4.67          & 33\%           & 0\%          & 15         \\ \hline
Others                                                   & 2.55    & 2.10    & 70\%   & 5\%   & 10          \\ \hline
\textit{Average}                                          & \textbf{6.33}  & \textbf{3.42} & \textbf{52\%} & \textbf{11\%} & \textbf{-} \\ \hline
\end{tabular}
\end{threeparttable}
}
\end{table}

\subsection{RQ4: Challenges in Bug Detection}\label{subsec:rq4}

\noindent\textbf{Methodology.} To detect a bug, developers first need to reproduce the symptoms reported in the Issues. Then, developers locate possible faulty code snippets based on the reported symptoms. Next, developers analyze the root cause of the bug and finally proceed to bug fixing. To find the challenge of detecting Jupyter bugs, we leverage four indicators adopted by previous studies \cite{Wang:2021An, Garcia:2020} on Issues to quantify the difficulty in detecting bugs.

\noindent$\bullet$ \textbf{Comments Num (\#Comment)}: The average number of comments under Issues. A high \#Comment indicates that a bug requires many discussions among users and developers.

\noindent$\bullet$  \textbf{Participants Num (\#Participant)}: The average number of participants under Issues. A high \#Participant indicates that a bug requires the combined efforts of multiple users and developers to be discovered and solved. 

\noindent$\bullet$  \textbf{Contributor Raised Issue Rate (\%Contributor)}: The rate of Issues raised by developers. A high rate means a bug is more likely to be detected by the developers, suggesting it is easier to detect the bug; and %It is obtained by:
%\frac{\# \text{developer-raised issuses}}{\# \text{all issues}},
%$$ 

\noindent$\bullet$  \textbf{External Issue Rate (\%External)}: The rate of Issues located in external repositories (e.g., PR in Jupytercore but the related Issues in Jupyterserver). A high rate suggests that the bug crosses different components. It indicates the complexity of the bug. % It is obtained by:
%$$
%\frac{\# \text{external repositories related issuses}}{\# \text{all Issues}},
%$$ 

In summary, \#Comment and \#Participant are positively related to the difficulty of detecting bugs. Furthermore, \%External is positively related to the complexity of bugs. \%Contributor is negatively related to the complexity of bugs. We adopt the non-dominated ranking~\cite{Mathew:2021,Ded:2002} to sort the difficulty in bug detection. Compared to sorting algorithms that aggregate objects using linear or nonlinear functions, non-dominated ranking orders multiple objects without aggregation (i.e., no weights need to be tuned). The only parameter to set in the non-dominated ranking is the optimal value of each indicator. To eliminate the effect of different magnitudes of indicators, we first normalize each indicator as follows: 

$$
\frac{indicator^i_j - indicator^i_{min} }{ indicator^i_{max} - indicator^i_{min}},
$$ 

where $1\leq i \leq 4$ and $1\leq j \leq 11$. For example, \#Comment of Launch in Table.~\ref{tab:challenge-symptom} is $indicator^1_4$. Thus, for positively related indicators, the optimal value is 1, and vice versa is 0. Non-dominated ranking sorts by calculating the euclidean distance from the normalized indicators to their optimal values.

\noindent\textbf{Results.} From the perspective of symptoms, SS-related bugs rank first in Table. ~\ref{tab:challenge-symptom}, with 8.25 for \#Comment, 4.75 for \#Participant, 25\% for Contributor, and 25\% for External of all SS Issues.  

\noindent$\blacktriangleright $\textbf{Challenge 1: Reproducing and locating SS-related bugs based on abstract symptom description. } By manually analyzing the Issues on SS, we find that half of the Issues on SS do not have the specific error message and reproduce method. Taking Notebook(Rep) Issue\#2503 ~\cite{notebook-issue-2503} as an example, a user requests that Jupyter should support binding to Unix sockets to avoid exposing IP and port information. Such security concern makes it difficult for developers to reproduce the bugs, especially when there is no error message reported. Thus, developers first attempt to reproduce the SS-related bugs based on their background knowledge.

After reproducing the SS-related bugs, developers need to locate the relative buggy code. However, it is hard to define the ``buggy code'' for an SS-related bug. Different from a traditional bug detection and fixing process, an SS-related bug may not have unique solutions. Normally, developers have to inspect all feasible solutions before determining the ``buggy code''. For example, in Notebook(Rep) Issue\#1074 ~\cite{notebook-issue-1074}, to avoid running the web browser in the root mode, developers propose different possible solutions:

\noindent$\bullet$ Solution (1): When a user opens the web browser in root mode, the web browser can send a warning message to the user;

\noindent$\bullet$ Solution (2): The kernel can deny users from running in root mode; and

\noindent$\bullet$ Solution (3): When the user launches the web browser, the front-end checks whether the Jupyter platform is running in root mode. If so, it can raise an assertion error and shut down the web browser.

Among these three possible solutions, (1) is considered to be ineffective and is abandoned by developers. (2) can restrict users from using IPykernel in many scenarios. For example, some users need to run the console in root mode in IPykernel. (3) successfully prevents users from running the web browser in root mode without any side effects. As a result, solution (3) is selected. Then, the developer locates buggy code snippets and fixes this Issue.

\begin{tcolorbox}[title=Finding 11,boxrule=1pt,boxsep=1pt,left=2pt,right=2pt,top=2pt,bottom=2pt]
According to the rank in Table. \ref{tab:challenge-symptom}, SS-related bugs are challenging to detect. Reproducing and locating SS-related bugs based on abstract symptom descriptions is a challenge for SS-related bugs because there can be multiple solutions for an SS-related bug. Normally, developers have to inspect all feasible solutions before determining the ``buggy code''.
\end{tcolorbox} 

As shown in Table. \ref{tab:challenge-symptom}, bugs related to Launch rank second. It is worth noting that Launch has the highest value of \#Comment and \#Participant, which suggests that detecting a Launch-related bug requires much effort among users and developers. By manually inspecting the conversations in the Issues and the commits in the pull requests, we conclude the second challenge in bug detection: Multiple root causes can lead to the same symptom.

\noindent$\blacktriangleright $
\textbf{Challenge 2: Multiple root causes can lead to the same symptom.} For some Launch-related bugs, we find that multiple root causes can lead to the same symptom. For example, in Notebook(Rep) Issue\#448, users encounter the error ''the Jupyter is not a command'' when launching the web browser. Two root causes can lead to this symptom. On the one hand, this symptom can be triggered by the Alg/Meth-related bug in the third-party library IJulia, which makes the conflicts between IJulia and Notebook(Rep). On the other hand, it can also be triggered by the Checking-related bug in Jupytercore, which leads to the Mac OSX does not add Jupyter to the system path when installing. These two root causes independently cause the same symptom. Developers can only solve this Issue completely when they find both root causes. Thus, it is challenging for developers to systematically find all root causes and detect bugs based on a single symptom.

\begin{tcolorbox}[title=Finding 12,boxrule=1pt,boxsep=1pt,left=2pt,right=2pt,top=2pt,bottom=2pt]
Launch ranks second in Table. \ref{tab:challenge-symptom}. The same symptom can be related to multiple root causes is the challenge for detecting Jupyter's bugs. Developers can only solve the Issue completely by finding all the root causes leading to the same symptom.
\end{tcolorbox}

\begin{table}[!htbp]
%\vspace{-1em}
\caption{Ranking of Root Causes for Bug Fixing}
%\vspace{-1em}
\label{tab:challenge-cause}
\scalebox{0.85}{
\begin{threeparttable}
    \begin{tabular}{lccccc}
\hline
\multicolumn{1}{c|}{\multirow{2}{*}{\textbf{Root Cause}}} & \multicolumn{5}{c}{\textbf{Pull Requests}}                                    \\ \cline{2-6} 
\multicolumn{1}{c|}{}                                     & \#Conversation & \#Commit      & LOCA           & LOCD           & Ranking    \\ \hline
Alg/Meth                      

& \textcolor{blue}{\textbf{8.46}}           & \textcolor{blue}{\textbf{3.32}}          & \textcolor{blue}{\textbf{76.26}}          & \textcolor{blue}{\textbf{27.08}}          & \textcolor{blue}{\textbf{1}}          \\ \hline

Ass/Ini                                                   & 4.90           & 1.71          & 11.15          & 7.39           & 8          \\ \hline
Checking                                                  & 4.62           & 1.81          & 21.63          & 4.46           & 7          \\ \hline
Logic                                                     & 5.89           & 1.86          & 25.57          & 7.71           & 6          \\ \hline
Data                                                      & 2.67           & 1.67          & 39.33          & 20.00          & 9          \\ \hline
Exter-API                                                 & 3.30           & 1.50          & 9.40           & 20.50          & 11         \\ \hline
Inter-API                                                 & 3.90           & 2.40          & 32.30          & 14.40          & 5          \\ \hline
Time/Perf                                                  & 3.75           & 3.75          & 32.75          & 9.5            & 4          \\ \hline
Config

& \textcolor{blue}{\textbf{4.8}}      & \textcolor{blue}{\textbf{2.88}}    & \textcolor{blue}{\textbf{21.65}}  & \textcolor{blue}{\textbf{41.05}}   & \textcolor{blue}{\textbf{2}}          \\ \hline

% & \textcolor{blue}{\textbf{4.84(4.8)}}      & \textcolor{blue}{\textbf{2.87(2.88)}}    & \textcolor{blue}{\textbf{171.05(21.65)}}  & \textcolor{blue}{\textbf{40.39(41.05)}}   & \textcolor{blue}{\textbf{2}}          \\ \hline
Non                                                       & 3.56           & 1.44          & 5.67           & 41.56          & 10         \\ \hline
Others                     
& 6.17 & 1.83    & 50.5  & 11.3   & 3          \\ \hline
\textit{Average}

%& 14.28(6.17)    & 2.86(1.83)    & 256.86(50.5)   & 251.86(11.3)   & 3          \\ \hline

& \textbf{5.71}  & \textbf{2.35} & \textbf{33.48} & \textbf{18.60} & \textbf{-} \\ \hline

%\textbf{5.86(5.71)}  & \textbf{2.37(2.35)} & \textbf{60.78(33.48)} & \textbf{22.89(18.60)} & \textbf{-} \\ \hline

\end{tabular}
\end{threeparttable}
}
%\begin{tablenotes}
  %  \small \item [1] Values in parentheses are calculated without PR 2066~\cite{notebook-pull-2066}.
    
  %  \small \item [2] Values in parentheses are calculated without PR 623~\cite{jupyterclient-pull-623}. 
    
  %  \small \item [3] Values in parentheses are calculated without PR 623~\cite{jupyterclient-pull-623} and 2066~\cite{notebook-pull-2066}
%\end{tablenotes}

\end{table}

\subsection{RQ5: Challenges in Bug Fixing}\label{subsec:rq5}

\noindent\textbf{Methodology.}
Bugs are mostly raised in Issues and fixed by developers in PRs. So to figure out the challenges in the bug fixing, we utilize the four indicators under PRs \cite{Wang:2021An, Garcia:2020} as follows:

\noindent$\bullet$ \textbf{Conversations Num (\#Conversation)}: The average number of conversations under PRs. A complex fix requires developers to work together to discuss plans and details. The higher the \#Conversation, the more difficult the bug is to fix;

\noindent$\bullet$ \textbf{Commits Num (\#Commit)}: The average number of commits under PRs. The higher the \#Commit, the more attempts it requires to fix the bug;

\noindent$\bullet$ \textbf{Lines of Code Added (LOCA)}: The average number of added lines of code under PRs. The higher the LOCA, the more effort is needed to modify the code; and

\noindent$\bullet$ \textbf{Lines of Code Deleted (LOCD)}: The average number of deleted lines of code under PRs. Similar to LOCA, the higher the LOCD, the more effort is needed to modify the code.

Therefore, all four indicators are positively related to the difficulty of bug fixing. Non-dominated ranking described in Sec.~\ref{subsec:rq4} is also employed here for ranking. Note that for LOCA and LOCD, we use their sum rather than independently in sorting. %We notice that the LOCA of Config and the LOCA and LOCD of Others are abnormally large in Table.~\ref{tab:challenge-cause}. By manual inspection, we find that the anomalies are caused by two unusual PRs. PR 2066~\cite{notebook-pull-2066} related to Config adds a new file "tools/patches/codemirror.js" containing 9124 LOCA to patch CodeMirror, which is a third-party library file for editing code~\cite{codeMirror}. PR 623~\cite{jupyterclient-pull-623} (related to Others) adds type annotations for lots of functions in many files to avoid potential bugs instead of fixing the existing bug. To maintain fairness in comparison and analysis, we recalculate the indicators without these two PRs. The results are recorded in parentheses in Table.~\ref{tab:challenge-cause} and \ref{tab:challenge-symptom}.

\noindent\textbf{Results.} 
Table.~\ref{tab:challenge-cause} suggests that bugs caused by Alg/Meth require the most effort to fix, which is consistent with finding 1 in Sec.~\ref{subsec:rq1}.

\noindent$\blacktriangleright $ \textbf{Challenge 1: Fixing one bug in a muti-module platform can introduce extra bugs.} Alg/Meth is the most difficult category of bugs to fix. This is especially the case for a muti-module platform, such as the Jupyter. The Jupyter platform is a complex software composed of many modules and functions. Fixing a bug in one function can interference the functionality of other relative modules in the Jupyter platform (e.g., incorrect output, hang, throw an error). As a result, such a fix can introduce new bugs to the Jupyter platform. Even worse, sometimes these new bugs can not be captured by running regression tests. It is because the executing traces of newly introduced bugs can not be covered by the regression test suites. To capture these new bugs, extra test cases are required. However, developers may not be aware of these new bugs, which makes the bug fixing process hard and challenging. It suggests that the Jupyter platform needs a more comprehensive and automated testing tool to help developers maintain their software efficiently.

\begin{tcolorbox}[title=Finding 13,boxrule=1pt,boxsep=1pt,left=2pt,right=2pt,top=2pt,bottom=2pt]
Alg/Meth is the most difficult category of bugs to fix in a muti-module platform like Jupyter. This is because fixing one bug can interference the functionality of other relative modules. As a result, extra bugs can be introduced. Even worse, some of them (i.e., extra bugs) cannot be captured with the regression test suite.
\end{tcolorbox}

%From the perspective of symptoms (Table. \ref{tab:challenge-symptom}), SS ranks first with values of 18.75, 6.25, 132.25, and 14.00, respectively. The most notable value is the \#Conversation, which is more than three times the others. It suggests that developers spend lots of time discussing the fixing solution. By analyzing the conversations in related PRs, we conclude that the second challenge:

%\noindent$\blacktriangleright $ \textbf{Challenge 2: Finding the appropriate solution for fixing security bugs.} A very appealing row in Table.~\ref{tab:challenge-symptom} is SS. Bugs related to SS occupy only a small part (i.e., 4 out of 388), but they are extremely difficult to fix. Bugs in this category can be linked to SSL errors, wrong permissions, and so forth. When encountering such bugs, participants often spend lots of time deciding the most appropriate solution. As described in Sec.~\ref{subsec:arch}, Jupyter platform has five components and each component has several modules. Where to implement fixes can minimize the impact on other components is of great concern to developers. This makes developers discuss a lot and make multiple attempts to solve bugs, which results in an extraordinarily high \#Conversation and \#Commit.

%\begin{tcolorbox}[title=Finding 15,boxrule=1pt,boxsep=1pt,left=2pt,right=2pt,top=2pt,bottom=2pt]
%From the perspective of symptoms, SS is a rare but most challenging category of bugs. Finding the appropriate %solution is a time-consuming process in fixing bugs.
%\end{tcolorbox} 

% 从Root cause角度进行发掘

% 从一个Root cause推出一个challenge

% 具体讲challenge

The ranking in Table.~\ref{tab:challenge-symptom} suggests that Config can be another root cause worth studying in-depth. Recall the finding 4 in Sec.~\ref{subsec:rq1}, bugs caused by Config often have a close connection with third-party libraries. 
% By analyzing PRs and Issues, we find that bugs related to third-party dependency library conflicts are also a major challenge in bug fixing.
%For bugs caused by upgrades or modifications of third-party libraries, developers have two basic solutions, including modifying the version of the library (e.g., using a compatible older version) and adapting the project's code to the new version. 

\noindent$\blacktriangleright $ \textbf{Challenge 2: Resolving third-party library conflicts in Config.}
% 冲突可能来自两方面：错误配置第三方库导致与当前库(e.g.,Notebook)产生冲突；两个第三方库产生冲突
A third-party library can conflict with its repository (e.g., Notebook(Rep)) or another third-party library in the Jupyter platform. 

% 对于前者，冲突源于版本号配置错误，这会导致各种各样不同的症状
The former conflict represents that the repository uses an incorrect version of the third-party library, which results in the modules in the repository working abnormally. This can lead to a variety of symptoms, such as front-end page rendering errors and user-entered code not displayed properly. The bug in Notebook(Rep) Issue\#3629 \cite{notebook-issue-3629} is an example. The user cannot open any notebook due to lack of the module \code{nbconvert.exporters.base}. Such module is introduced in \code{nbconvert} 5.0.0~\cite{nbconvert} library, but Jupyter platform fails to set the minimum version of \code{nbconvert}, which make users with outdated \code{nbconvert} encounter this bug.

% 第一种冲突的例子(错误配置第三方库导致与当前库(e.g.,Notebook)产生冲突)

% 对于后者，单独使用某个库A没有问题，但是一同使用库A和库B就会需要额外配置，否则会造成bug
% 因为这两个库的某些功能模块重叠或是因为使用了共享资源
% 和前者一样，症状也是多种多样的，开发者很难区分两者，即很难找到到底是哪种冲突。

% 第二种冲突的例子，两个库都提供了codemirror.js导致Jupyter加载错误。
% 负责修复的开发者不知道是由两个库共同导致了这一问题，故而一直试图找codemirror的配置错误，直到另一位开发者提醒在没有引入webpack前一切工作正常，才最终找到是由两个库共同造成这一bug。

The latter conflict arises because both libraries contain files, modules, or executables with the same name but different contents. Using the incorrect files or modules can lead to errors. Taking Notebook(Rep) PR\#1547~\cite{notebook-issue-1547} as an example, the root of the bug is the existence of two \code{codemirror.js}. One is imported by the developer and another one is bundled by Webpack~\cite{webpack}. The Jupyter platform uses the bundled version, which results in the bug. 

Currently, developers resolve dependency conflicts only by following a guess-and-check strategy. Specifically, developers first presume the potential problematic third-party library based on which modules work incorrectly. Next, they manually check the relevant code to verify whether this library is the culprit of the bug. This process is time-consuming and often requires multiple developers to work together to figure out the source of the conflict. A tool that can automatically detect which dependency libraries conflict can help developers fix these bugs more efficiently. 

\begin{tcolorbox}[title=Finding 14,boxrule=1pt,boxsep=1pt,left=2pt,right=2pt,top=2pt,bottom=2pt]
Config bugs are the second most difficult to fix, with the main challenge being identifying and resolving third-party dependency library conflicts. Incorporating an automated dependency conflict monitoring tool into Jupyter can help developers overcome this challenge.
\end{tcolorbox}

\section{Discussion}\label{sec:discussion}
\subsection{Implications for Further Research}\label{sec:suggestions}
In September 2018, more than 2.5 million Jupyter notebook repositories are hosted on GitHub, which are 10 times more than that in 2015 \cite{rule2018}. Ensuring the robustness of the Jupyter platform not only benefits the Jupyter developers but also benefits the users' experience of the Jupyter platform. Thus, we propose the following suggestions for stakeholders.

\noindent\textbf{For Jupyter developers,} the suggestions are:

% Be careful to variable Ass/Ini in front-end programming finding 2,9
\noindent $\bullet$ Finding 2 and 9 suggest that developers should be careful when editing the front-end UI attributes. Mistakes in these attributes can lead to DGUI errors. Such mistakes can be easily overlooked.

% Be careful Config finding 4 15
\noindent $\bullet$ Finding 4 and 14 suggest that developers should continuously pay attention to configurations, especially for configuring third-party dependencies. Furthermore, downstream components (e.g., Notebook(Rep)) can be more likely to have Config bugs and dependency conflicts, which can be hard to detect.

% developer should  consider all possible root causes even meet the same symptom finding 12
\noindent $\bullet$ Finding 12 reveals that multiple root causes can lead to the same symptom. It suggests that developers should enhance the debugging and logging systems inside the Jupyter platform. Such enhancement can assist developers in exploring and fixing all bugs that can lead to a symptom.

\noindent\textbf{For tool builders and researchers,} the suggestions are:
% 针对 finding 13 对于Jupyter这样多组件多模块的大型软件 开发者在修复bug可能会无意识地引入新的bug，
% 老的testcase无法检测新引入的bug 需要能够根据代码修改自动生成新test cases的测试工具

\noindent $\bullet$ \underline{Automation Testing Tools for Repairs}: Finding 13 implies that some bug fixes can introduce extra bugs to the Jupyter platform as the executing traces of some extra bugs can be excluded from the current regression test suites in the Jupyter. Thus, tool builders are supposed to build test case generators based on repairs. 

\noindent $\bullet$ \underline{Detecting and Fixing Tools for Dependency Conflicts}: 
% 对于finding 14 jupyter拥有大量的第三方依赖库
Jupyter has a great number of third-party dependency libraries (e.g., CodeMirror~\cite{codeMirror}, Tornado ~\cite{tornado}).
% 依赖库冲突bug 在jupyter中很常见，开发者目前只能通过手动排查的方式来确定冲突来源(i.e., guess-and-check)，这种方式耗时且不可靠
According to finding 14, dependency conflicts bugs are plentiful in Config bugs. Developers can currently only determine the source of conflicts with manual checking, which is time-consuming and unreliable.
% 故而，工具开发者应该开发对应的依赖库冲突检测与修复工具来帮助Jupyter的维护。
Therefore, tool builders can help maintain Jupyter by developing tools for detecting and fixing dependency conflicts.

\noindent$\bullet$ \underline{Tools for detecting front-end page rendering errors:} Finding 2 and 9 suggest that errors in setting attributes for front-end UI elements can lead to various page rendering errors. Such errors can be easily overlooked. Thus, developers can build tools to detect such page rendering errors. %For example, object detecting algorithms can be used to detect UI elements and checks if UI elements are overlapping.

\subsection{Threats to Validity}

\noindent\textbf{Threats to internal validity.}
The most dominant internal validity stems from deviations and errors in bug classification. To minimize bias, we follow the data collection procedure presented in an ICSE'20 paper~\cite{Garcia:2020}. To ensure that we can focus on the real bugs and their fixes, we identify closed and merged PRs with a list of words that have the meaning of bugs, as stated in detail in sec.~\ref{sec:classify-label}. Finally, we manually checked all PRs and their related issues to filter out the irrelevant parts. To ensure the reliability of the bug classification results, two authors analyze the bugs separately and discuss the discrepancy with another author until reaching unification. We adopt the taxonomy used in the existing research~\cite{Garcia:2020, Seaman:2008, Thung:2012} to label the bugs. Through these efforts, we mitigate the threats to internal validity to the greatest extent possible.

\noindent\textbf{Threats to external validity.}
We select the closed and merged pull requests with their relevant issues until March 15, 2022. The opened pull requests and their related issues are not counted in statistics, making us miss some bugs that are continuously tracked. Also, the result of the symptoms can deviate from facts because some issues may have been updated recently. Therefore, readers should be cautious when applying our findings.

%in Config and Others from the statistics. 

\section{Related Work}\label{sec:related-work}
Despite Jupyter's popularity, research works on Jupyter and Jupyter notebooks are still limited. The existing research can be grouped into the following categories:

\noindent\textbf{Code Reproducibility and Reuse on Jupyter notebooks.} Pimentel et al. conducted a large-scale study on the quality and reproducibility of Jupyter notebooks \cite{Pimentel:2019}. They studied 1.4 million notebooks from GitHub, showing that only 24.11\% of Jupyter notebooks can be executed without exception, and only around 4\% of notebooks can be reproduced with the same results. Koenzen et al. \cite{Koenzen:2020} explored the way of code duplications in Jupyter notebooks and identified the potential barriers to code reuse. Besides, Wang et al. \cite{Wang:2020ASE} first studied whether existing notebooks can be executed successfully (i.e., reproducibility). Then, they proposed a prototype named Osiris, which takes a notebook as an input and outputs the possible execution schemes to reproduce the notebook.

%They found that the repositories in their sample have a mean self-duplication rate of around 7\%. Developers prefer to find reuse code online. Besides, Wang et al \cite{Wang:2020ASE} first studied whether execting notebooks can be executed successfully (i.e. reproducibility). Then, they proposed a prototype named Osris, which takes a notebook as an input and outputs the possible execution schemes to reproduce the notebook.

\noindent\textbf{Restore/Repair and Code Quality on Jupyter notebook.} Zhu et al. \cite{Zhu:2021} proposed RELANCER, which is an automatic technique that can restore the executability of broken Jupyter notebooks by upgrading deprecated APIs. Wang et al. \cite{Wang:2021} developed SnifferDog to restore the execution environments for executing Jupyter notebooks. Specifically, SnifferDog first collects the APIs of Python packages to build the database and then analyzes the notebooks to determine the candidate packages and versions. Wang et al. 's work \cite{Wang:2020} conducted a preliminary study on code quality for Jupyter notebooks. They found that the existing notes are with poor quality codes, which requires quality control on Jupyter notebooks.

\noindent\textbf{Empirical Study on Software Bugs.} Franco et al. \cite{Franco:2017} classified 269 numerical bugs from five famous numerical libraries and further analyzed their occurrence frequency, symptoms, and fixes. Romano et al. \cite{Romano:2021} analyzed 1054 bugs in three widely-used open-source WebAssembly compilers, including AssemblyScript, Emscripten, and Rustc/Wasm-Bindgen. Garcia et al. \cite{Garcia:2020} categorized the root cause and symptom of 499 bugs in two representative autonomous vehicle software (i.e., Apollo and Autoware). Shen et al. \cite{Shen:2021} classified the root cause, symptom, and occurrence stage of 603 bugs in three popular deep learning (DL) compilers (i.e., TVM, Glow, and nGraph). Gao et al. \cite{Gao:2018} analyzed and classified 103 crash recovery bugs in four open-source distributed systems. Eghbali et al. \cite{Eghbali:2020} conducted a comprehensive analysis of 204 string-related bugs in a dataset containing 13 top-starred JavaScript projects on GitHub.

In summary, the existing research works are more focused on restoring Jupyter notebooks' environment and making code reproducible. However, the bugs on the Jupyter are not researched. Our study fulfills this gap and assists Jupyter developers in bug detection and fixing on the Jupyter.

\section{Conclusion and Future Work}\label{sec:conclusion}
Given the fact that Jupyter is the most widely used platform for developing Jupyter notebooks, it is critical that software-engineering researchers build a robust and secure Jupyter platform. In this paper, we conduct a systematic study on bugs for the Jupyter platform. We identify 11 root causes and 11 symptoms across five components in the Jupyter platform from 387 Jupyter bugs. Both researchers and developers can benefit from this study. For developers, we summarize 14 findings to help them deal with bugs. For researchers, we distill the challenges which require additional research effort (see Sec.\ref{sec:suggestions}). In the future, we plan to propose techniques 
for automatic test suite generation for the Jupyter platform, especially for testing program repairs. Additionally, we plan to refine our study's results to build a benchmark dataset for automatic program repair techiques for the Jupyter platform. 

\section{Data Availability}
The results and data can be found at: \cite{artifact}.

%\newpage

\bibliographystyle{ACM-Reference-Format}
\balance
\bibliography{reference}


\begin{thebibliography}{99}

%%% the Web address.

\ifx \showCODEN    \undefined \def \showCODEN     #1{\unskip}     \fi
\ifx \showDOI      \undefined \def \showDOI       #1{#1}\fi
\ifx \showISBNx    \undefined \def \showISBNx     #1{\unskip}     \fi
\ifx \showISBNxiii \undefined \def \showISBNxiii  #1{\unskip}     \fi
\ifx \showISSN     \undefined \def \showISSN      #1{\unskip}     \fi
\ifx \showLCCN     \undefined \def \showLCCN      #1{\unskip}     \fi
\ifx \shownote     \undefined \def \shownote      #1{#1}          \fi
\ifx \showarticletitle \undefined \def \showarticletitle #1{#1}   \fi
\ifx \showURL      \undefined \def \showURL       {\relax}        \fi
% invisible to TeX
\providecommand\bibfield[2]{#2}
\providecommand\bibinfo[2]{#2}
\providecommand\natexlab[1]{#1}
\providecommand\showeprint[2][]{arXiv:#2}

\bibitem[Anyio(2022)]%
        {anyio}
\bibfield{author}{\bibinfo{person}{Anyio}.} \bibinfo{year}{2022}\natexlab{}.
\newblock \bibinfo{title}{AnyIO: an asynchronous networking and concurrency
  library}.
\newblock
\newblock
\newblock
\shownote{\url{https://github.com/agronholm/anyio}}.


\bibitem[Artifact(2022)]%
        {artifact}
\bibfield{author}{\bibinfo{person}{Online Artifact}.}
  \bibinfo{year}{2022}\natexlab{}.
\newblock \bibinfo{title}{Online Artifact}.
\newblock
\newblock
\newblock
\shownote{\url{https://sites.google.com/view/jupyter-bugs/}}.


\bibitem[Ball et~al\mbox{.}(2003)]%
        {Thomas:2003}
\bibfield{author}{\bibinfo{person}{Thomas Ball}, \bibinfo{person}{Mayur Naik},
  {and} \bibinfo{person}{Sriram~K. Rajamani}.} \bibinfo{year}{2003}\natexlab{}.
\newblock \showarticletitle{From symptom to cause: localizing errors in
  counterexample traces}. In \bibinfo{booktitle}{\emph{Proceedings of POPL}}.
  \bibinfo{pages}{97--105}.
\newblock


\bibitem[Barba(2021)]%
        {Lorena:2021}
\bibfield{author}{\bibinfo{person}{Lorena~A. Barba}.}
  \bibinfo{year}{2021}\natexlab{}.
\newblock \showarticletitle{The Python/Jupyter Ecosystem: Today's
  Problem-Solving Environment for Computational Science}.
\newblock \bibinfo{journal}{\emph{Computing in Science \& Engineering}}
  \bibinfo{volume}{23}, \bibinfo{number}{3} (\bibinfo{year}{2021}),
  \bibinfo{pages}{5--9}.
\newblock


\bibitem[Beg et~al\mbox{.}(2021)]%
        {Marijan:2021}
\bibfield{author}{\bibinfo{person}{Marijan Beg}, \bibinfo{person}{Juliette
  Taka}, \bibinfo{person}{Thomas Kluyver}, \bibinfo{person}{Alexander
  Konovalov}, \bibinfo{person}{Min Ragan{-}Kelly}, \bibinfo{person}{Nicolas~M.
  Thi{\'{e}}ry}, {and} \bibinfo{person}{Hans Fangohr}.}
  \bibinfo{year}{2021}\natexlab{}.
\newblock \showarticletitle{Using Jupyter for Reproducible Scientific
  Workflows}.
\newblock \bibinfo{journal}{\emph{Computing in Science \& Engineering}}
  \bibinfo{volume}{23}, \bibinfo{number}{2} (\bibinfo{year}{2021}),
  \bibinfo{pages}{36--46}.
\newblock


\bibitem[codemirror(2022)]%
        {codeMirror}
\bibfield{author}{\bibinfo{person}{codemirror}.}
  \bibinfo{year}{2022}\natexlab{}.
\newblock \bibinfo{title}{CodeMirror}.
\newblock
\newblock
\newblock
\shownote{\url{https://github.com/codemirror/CodeMirror}}.


\bibitem[Deb et~al\mbox{.}(2002)]%
        {Ded:2002}
\bibfield{author}{\bibinfo{person}{Kalyanmoy Deb}, \bibinfo{person}{Samir
  Agrawal}, \bibinfo{person}{Amrit Pratap}, {and} \bibinfo{person}{T.
  Meyarivan}.} \bibinfo{year}{2002}\natexlab{}.
\newblock \showarticletitle{A fast and elitist multiobjective genetic
  algorithm: {NSGA-II}}.
\newblock \bibinfo{journal}{\emph{{IEEE} Trans. Evol. Comput.}}
  \bibinfo{volume}{6}, \bibinfo{number}{2} (\bibinfo{year}{2002}),
  \bibinfo{pages}{182--197}.
\newblock


\bibitem[Di~Franco et~al\mbox{.}(2017)]%
        {Franco:2017}
\bibfield{author}{\bibinfo{person}{Anthony Di~Franco}, \bibinfo{person}{Hui
  Guo}, {and} \bibinfo{person}{Cindy Rubio-González}.}
  \bibinfo{year}{2017}\natexlab{}.
\newblock \showarticletitle{A comprehensive study of real-world numerical bug
  characteristics}. In \bibinfo{booktitle}{\emph{Proceedings of ASE}}.
  \bibinfo{pages}{509--519}.
\newblock


\bibitem[Eghbali and Pradel(2020)]%
        {Eghbali:2020}
\bibfield{author}{\bibinfo{person}{Aryaz Eghbali} {and}
  \bibinfo{person}{Michael Pradel}.} \bibinfo{year}{2020}\natexlab{}.
\newblock \showarticletitle{No Strings Attached: An Empirical Study of
  String-related Software Bugs}. In \bibinfo{booktitle}{\emph{Proceedings of
  ASE}}. \bibinfo{pages}{956--967}.
\newblock


\bibitem[Fangohr et~al\mbox{.}(2021)]%
        {Hans:2021}
\bibfield{author}{\bibinfo{person}{Hans Fangohr}, \bibinfo{person}{Thomas
  Kluyver}, {and} \bibinfo{person}{Massimo DiPierro}.}
  \bibinfo{year}{2021}\natexlab{}.
\newblock \showarticletitle{Jupyter in Computational Science}.
\newblock \bibinfo{journal}{\emph{Computing in Science \& Engineering}}
  \bibinfo{volume}{23}, \bibinfo{number}{2} (\bibinfo{year}{2021}),
  \bibinfo{pages}{5--6}.
\newblock


\bibitem[for Standardization(ISO)(2022)]%
        {iso8601}
\bibfield{author}{\bibinfo{person}{International~Organization for
  Standardization(ISO)}.} \bibinfo{year}{2022}\natexlab{}.
\newblock \bibinfo{title}{ISO8601: Date and time format}.
\newblock
\newblock
\newblock
\shownote{\url{https://www.iso.org/iso-8601-date-and-time-format.html}}.


\bibitem[Gao et~al\mbox{.}(2018)]%
        {Gao:2018}
\bibfield{author}{\bibinfo{person}{Yu Gao}, \bibinfo{person}{Wensheng Dou},
  \bibinfo{person}{Feng Qin}, \bibinfo{person}{Chushu Gao},
  \bibinfo{person}{Dong Wang}, \bibinfo{person}{Jun Wei},
  \bibinfo{person}{Ruirui Huang}, \bibinfo{person}{Li Zhou}, {and}
  \bibinfo{person}{Yongming Wu}.} \bibinfo{year}{2018}\natexlab{}.
\newblock \showarticletitle{An empirical study on crash recovery bugs in
  large-scale distributed systems}. In \bibinfo{booktitle}{\emph{Proceedings of
  ESEC/FSE}}, \bibfield{editor}{\bibinfo{person}{Gary~T. Leavens},
  \bibinfo{person}{Alessandro Garcia}, {and} \bibinfo{person}{Corina~S.
  Pasareanu}} (Eds.). \bibinfo{pages}{539--550}.
\newblock


\bibitem[Garcia et~al\mbox{.}(2020)]%
        {Garcia:2020}
\bibfield{author}{\bibinfo{person}{Joshua Garcia}, \bibinfo{person}{Yang Feng},
  \bibinfo{person}{Junjie Shen}, \bibinfo{person}{Sumaya Almanee},
  \bibinfo{person}{Yuan Xia}, {and} \bibinfo{person}{Qi~Alfred Chen}.}
  \bibinfo{year}{2020}\natexlab{}.
\newblock \showarticletitle{A Comprehensive Study of Autonomous Vehicle Bugs}.
  In \bibinfo{booktitle}{\emph{Proceedings of ICSE}}.
  \bibinfo{pages}{385–396}.
\newblock


\bibitem[Granger and P{\'{e}}rez(2021)]%
        {Brian:2021}
\bibfield{author}{\bibinfo{person}{Brian~E. Granger} {and}
  \bibinfo{person}{Fernando P{\'{e}}rez}.} \bibinfo{year}{2021}\natexlab{}.
\newblock \showarticletitle{Jupyter: Thinking and Storytelling With Code and
  Data}.
\newblock \bibinfo{journal}{\emph{Computing in Science \& Engineering}}
  \bibinfo{volume}{23}, \bibinfo{number}{2} (\bibinfo{year}{2021}),
  \bibinfo{pages}{7--14}.
\newblock


\bibitem[IJuliaKernel(2022)]%
        {IJuliaKernel}
\bibfield{author}{\bibinfo{person}{IJuliaKernel}.}
  \bibinfo{year}{2022}\natexlab{}.
\newblock \bibinfo{title}{IJuliaKernel: Julia kernel for Jupyter}.
\newblock
\newblock
\newblock
\shownote{\url{https://juliapackages.com/p/ijulia}}.


\bibitem[IPykernel(2022)]%
        {ipykernel}
\bibfield{author}{\bibinfo{person}{IPykernel}.}
  \bibinfo{year}{2022}\natexlab{}.
\newblock \bibinfo{title}{IPykernel}.
\newblock
\newblock
\newblock
\shownote{\url{https://github.com/ipython/ipykernel}}.


\bibitem[IPython(2022)]%
        {ipyparallel}
\bibfield{author}{\bibinfo{person}{IPython}.} \bibinfo{year}{2022}\natexlab{}.
\newblock \bibinfo{title}{IPython Parallel}.
\newblock
\newblock
\newblock
\shownote{\url{https://github.com/ipython/ipyparallel}}.


\bibitem[IRkernel(2022)]%
        {IRkernel}
\bibfield{author}{\bibinfo{person}{IRkernel}.} \bibinfo{year}{2022}\natexlab{}.
\newblock \bibinfo{title}{IRkernel: R kernel for Jupyter}.
\newblock
\newblock
\newblock
\shownote{\url{https://github.com/IRkernel/IRkernel}}.


\bibitem[Islam et~al\mbox{.}(2019)]%
        {Islam:2019}
\bibfield{author}{\bibinfo{person}{Md~Johirul Islam}, \bibinfo{person}{Giang
  Nguyen}, \bibinfo{person}{Rangeet Pan}, {and} \bibinfo{person}{Hridesh
  Rajan}.} \bibinfo{year}{2019}\natexlab{}.
\newblock \showarticletitle{A Comprehensive Study on Deep Learning Bug
  Characteristics}. In \bibinfo{booktitle}{\emph{Proceedings of ESEC/FSE}}.
  \bibinfo{pages}{510–520}.
\newblock


\bibitem[Jupyter(2015a)]%
        {ipykernel-pull-18}
\bibfield{author}{\bibinfo{person}{Jupyter}.} \bibinfo{year}{2015}\natexlab{a}.
\newblock \bibinfo{title}{Ipykernel pull\#18}.
\newblock
\newblock
\newblock
\shownote{\url{https://github.com/ipython/ipykernel/pull/18}}.


\bibitem[Jupyter(2015b)]%
        {notebook-issue-792}
\bibfield{author}{\bibinfo{person}{Jupyter}.} \bibinfo{year}{2015}\natexlab{b}.
\newblock \bibinfo{title}{Jupyter Notebook Issue\#792}.
\newblock
\newblock
\newblock
\shownote{\url{https://github.com/jupyter/notebook/issues/792}}.


\bibitem[Jupyter(2015c)]%
        {notebook-issue-96}
\bibfield{author}{\bibinfo{person}{Jupyter}.} \bibinfo{year}{2015}\natexlab{c}.
\newblock \bibinfo{title}{Jupyter Notebook Issue\#96}.
\newblock
\newblock
\newblock
\shownote{\url{https://github.com/jupyter/notebook/issues/96}}.


\bibitem[Jupyter(2015d)]%
        {notebook-pull-22}
\bibfield{author}{\bibinfo{person}{Jupyter}.} \bibinfo{year}{2015}\natexlab{d}.
\newblock \bibinfo{title}{Jupyter Notebook PR\#22}.
\newblock
\newblock
\newblock
\shownote{\url{https://github.com/jupyter/notebook/pull/22}}.


\bibitem[Jupyter(2015e)]%
        {notebook-issue-52}
\bibfield{author}{\bibinfo{person}{Jupyter}.} \bibinfo{year}{2015}\natexlab{e}.
\newblock \bibinfo{title}{Notebook issue\#52}.
\newblock
\newblock
\newblock
\shownote{\url{https://github.com/jupyter/notebook/issues/52}}.


\bibitem[Jupyter(2015f)]%
        {notebook-pull-799}
\bibfield{author}{\bibinfo{person}{Jupyter}.} \bibinfo{year}{2015}\natexlab{f}.
\newblock \bibinfo{title}{Notebook pull\#799}.
\newblock
\newblock
\newblock
\shownote{\url{https://github.com/jupyter/notebook/pull/799}}.


\bibitem[Jupyter(2016a)]%
        {notebook-issue-1003}
\bibfield{author}{\bibinfo{person}{Jupyter}.} \bibinfo{year}{2016}\natexlab{a}.
\newblock \bibinfo{title}{Jupyter Notebook Issue\#1003}.
\newblock
\newblock
\newblock
\shownote{\url{https://github.com/jupyter/notebook/issues/1003}}.


\bibitem[Jupyter(2016b)]%
        {notebook-pull-1011}
\bibfield{author}{\bibinfo{person}{Jupyter}.} \bibinfo{year}{2016}\natexlab{b}.
\newblock \bibinfo{title}{Jupyter Notebook PR\#1011}.
\newblock
\newblock
\newblock
\shownote{\url{https://github.com/jupyter/notebook/pull/1011}}.


\bibitem[Jupyter(2016c)]%
        {notebook-issue-1074}
\bibfield{author}{\bibinfo{person}{Jupyter}.} \bibinfo{year}{2016}\natexlab{c}.
\newblock \bibinfo{title}{Notebook issue\#1074}.
\newblock
\newblock
\newblock
\shownote{\url{https://github.com/jupyter/notebook/issues/1074}}.


\bibitem[Jupyter(2016d)]%
        {notebook-issue-1547}
\bibfield{author}{\bibinfo{person}{Jupyter}.} \bibinfo{year}{2016}\natexlab{d}.
\newblock \bibinfo{title}{Notebook issue\#1547}.
\newblock
\newblock
\newblock
\shownote{\url{https://github.com/jupyter/notebook/issues/1547}}.


\bibitem[Jupyter(2016e)]%
        {notebook-issue-1977}
\bibfield{author}{\bibinfo{person}{Jupyter}.} \bibinfo{year}{2016}\natexlab{e}.
\newblock \bibinfo{title}{Notebook issue\#1977}.
\newblock
\newblock
\newblock
\shownote{\url{https://github.com/jupyter/notebook/issues/1977}}.


\bibitem[Jupyter(2016f)]%
        {notebook-pull-1652}
\bibfield{author}{\bibinfo{person}{Jupyter}.} \bibinfo{year}{2016}\natexlab{f}.
\newblock \bibinfo{title}{Notebook pull\#1652}.
\newblock
\newblock
\newblock
\shownote{\url{https://github.com/jupyter/notebook/pull/1652}}.


\bibitem[Jupyter(2017a)]%
        {notebook-issue-2203}
\bibfield{author}{\bibinfo{person}{Jupyter}.} \bibinfo{year}{2017}\natexlab{a}.
\newblock \bibinfo{title}{Jupyter Notebook Issue\#2203}.
\newblock
\newblock
\newblock
\shownote{\url{https://github.com/jupyter/notebook/issues/2203}}.


\bibitem[Jupyter(2017b)]%
        {jupyterclient-pull-254}
\bibfield{author}{\bibinfo{person}{Jupyter}.} \bibinfo{year}{2017}\natexlab{b}.
\newblock \bibinfo{title}{Jupyterclient pull\#254}.
\newblock
\newblock
\newblock
\shownote{\url{https://github.com/jupyter/jupyter_client/pull/254}}.


\bibitem[Jupyter(2017c)]%
        {notebook-issue-2503}
\bibfield{author}{\bibinfo{person}{Jupyter}.} \bibinfo{year}{2017}\natexlab{c}.
\newblock \bibinfo{title}{Notebook issue\#2503}.
\newblock
\newblock
\newblock
\shownote{\url{https://github.com/jupyter/notebook/issues/2503}}.


\bibitem[Jupyter(2017d)]%
        {notebook-pull-2159}
\bibfield{author}{\bibinfo{person}{Jupyter}.} \bibinfo{year}{2017}\natexlab{d}.
\newblock \bibinfo{title}{Notebook pull\#2159}.
\newblock
\newblock
\newblock
\shownote{\url{https://github.com/jupyter/notebook/pull/2159}}.


\bibitem[Jupyter(2018a)]%
        {notebook-issue-3629}
\bibfield{author}{\bibinfo{person}{Jupyter}.} \bibinfo{year}{2018}\natexlab{a}.
\newblock \bibinfo{title}{Notebook issue\#3629}.
\newblock
\newblock
\newblock
\shownote{\url{https://github.com/jupyter/notebook/issues/3629}}.


\bibitem[Jupyter(2018b)]%
        {notebook-pull-4236}
\bibfield{author}{\bibinfo{person}{Jupyter}.} \bibinfo{year}{2018}\natexlab{b}.
\newblock \bibinfo{title}{Notebook pull\#4236}.
\newblock
\newblock
\newblock
\shownote{\url{https://github.com/jupyter/notebook/pull/4236}}.


\bibitem[Jupyter(2019a)]%
        {ipykernel-pull-390}
\bibfield{author}{\bibinfo{person}{Jupyter}.} \bibinfo{year}{2019}\natexlab{a}.
\newblock \bibinfo{title}{Ipykernel pull\#390}.
\newblock
\newblock
\newblock
\shownote{\url{https://github.com/ipython/ipykernel/pull/390}}.


\bibitem[Jupyter(2019b)]%
        {jupyterserver-issue-42}
\bibfield{author}{\bibinfo{person}{Jupyter}.} \bibinfo{year}{2019}\natexlab{b}.
\newblock \bibinfo{title}{Jupyterserver Issue\#42}.
\newblock
\newblock
\newblock
\shownote{\url{https://github.com/jupyter-server/jupyter_server/issues/42}}.


\bibitem[Jupyter(2020a)]%
        {notebook-issue-5190}
\bibfield{author}{\bibinfo{person}{Jupyter}.} \bibinfo{year}{2020}\natexlab{a}.
\newblock \bibinfo{title}{Jupyter Notebook Issue\#5190}.
\newblock
\newblock
\newblock
\shownote{\url{https://github.com/jupyter/notebook/issues/5190}}.


\bibitem[Jupyter(2020b)]%
        {notebook-issue-5502}
\bibfield{author}{\bibinfo{person}{Jupyter}.} \bibinfo{year}{2020}\natexlab{b}.
\newblock \bibinfo{title}{Jupyter Notebook Issue\#5502}.
\newblock
\newblock
\newblock
\shownote{\url{https://github.com/jupyter/notebook/issues/5502}}.


\bibitem[Jupyter(2020c)]%
        {jupyterclient-issue-591}
\bibfield{author}{\bibinfo{person}{Jupyter}.} \bibinfo{year}{2020}\natexlab{c}.
\newblock \bibinfo{title}{Jupyterclient Issue\#591}.
\newblock
\newblock
\newblock
\shownote{\url{https://github.com/jupyter/jupyter_client/issues/591}}.


\bibitem[Jupyter(2020d)]%
        {jupytercore-pull-183}
\bibfield{author}{\bibinfo{person}{Jupyter}.} \bibinfo{year}{2020}\natexlab{d}.
\newblock \bibinfo{title}{Jupytercore PR\#183}.
\newblock
\newblock
\newblock
\shownote{\url{https://github.com/jupyter/jupyter_core/pull/183}}.


\bibitem[Jupyter(2020e)]%
        {notebook-pull-5136}
\bibfield{author}{\bibinfo{person}{Jupyter}.} \bibinfo{year}{2020}\natexlab{e}.
\newblock \bibinfo{title}{Notebook pull\#5136}.
\newblock
\newblock
\newblock
\shownote{\url{https://github.com/jupyter/notebook/pull/5136}}.


\bibitem[Jupyter(2021a)]%
        {ipykernel-issue-694}
\bibfield{author}{\bibinfo{person}{Jupyter}.} \bibinfo{year}{2021}\natexlab{a}.
\newblock \bibinfo{title}{Ipykernel Issue\#694}.
\newblock
\newblock
\newblock
\shownote{\url{https://github.com/ipython/ipykernel/issues/694}}.


\bibitem[Jupyter(2021b)]%
        {ipykernel-issue-742}
\bibfield{author}{\bibinfo{person}{Jupyter}.} \bibinfo{year}{2021}\natexlab{b}.
\newblock \bibinfo{title}{Ipykernel Issue\#742}.
\newblock
\newblock
\newblock
\shownote{\url{https://github.com/ipython/ipykernel/issues/742}}.


\bibitem[Jupyter(2021c)]%
        {jupyterclient-pull-607}
\bibfield{author}{\bibinfo{person}{Jupyter}.} \bibinfo{year}{2021}\natexlab{c}.
\newblock \bibinfo{title}{Jupyterclient pull\#607}.
\newblock
\newblock
\newblock
\shownote{\url{https://github.com/jupyter/jupyter_client/pull/607}}.


\bibitem[Jupyter(2021d)]%
        {jupyterclient-pull-703}
\bibfield{author}{\bibinfo{person}{Jupyter}.} \bibinfo{year}{2021}\natexlab{d}.
\newblock \bibinfo{title}{Jupyterclient pull\#703}.
\newblock
\newblock
\newblock
\shownote{\url{https://github.com/jupyter/jupyter_client/pull/703}}.


\bibitem[Jupyter(2021e)]%
        {jupyterclient-pull-717}
\bibfield{author}{\bibinfo{person}{Jupyter}.} \bibinfo{year}{2021}\natexlab{e}.
\newblock \bibinfo{title}{Jupyterclient pull\#717}.
\newblock
\newblock
\newblock
\shownote{\url{https://github.com/jupyter/jupyter_client/pull/717}}.


\bibitem[Jupyter(2021f)]%
        {jupyterserver-issue-591}
\bibfield{author}{\bibinfo{person}{Jupyter}.} \bibinfo{year}{2021}\natexlab{f}.
\newblock \bibinfo{title}{Jupyterserver Issue\#591}.
\newblock
\newblock
\newblock
\shownote{\url{https://github.com/jupyter-server/jupyter_server/issues/591}}.


\bibitem[Jupyter(2021g)]%
        {jupyterserver-pull-380}
\bibfield{author}{\bibinfo{person}{Jupyter}.} \bibinfo{year}{2021}\natexlab{g}.
\newblock \bibinfo{title}{Jupyterserver pull\#380}.
\newblock
\newblock
\newblock
\shownote{\url{https://github.com/jupyter-server/jupyter_server/pull/380}}.


\bibitem[Jupyter(2021h)]%
        {jupyterserver-pull-473}
\bibfield{author}{\bibinfo{person}{Jupyter}.} \bibinfo{year}{2021}\natexlab{h}.
\newblock \bibinfo{title}{Jupyterserver pull\#473}.
\newblock
\newblock
\newblock
\shownote{\url{https://github.com/jupyter-server/jupyter_server/pull/473}}.


\bibitem[Jupyter(2021i)]%
        {jupyterserver-pull-482}
\bibfield{author}{\bibinfo{person}{Jupyter}.} \bibinfo{year}{2021}\natexlab{i}.
\newblock \bibinfo{title}{Jupyterserver pull\#482}.
\newblock
\newblock
\newblock
\shownote{\url{https://github.com/jupyter-server/jupyter_server/pull/482}}.


\bibitem[Jupyter(2021j)]%
        {jupyterserver-pull-521}
\bibfield{author}{\bibinfo{person}{Jupyter}.} \bibinfo{year}{2021}\natexlab{j}.
\newblock \bibinfo{title}{Jupyterserver pull\#521}.
\newblock
\newblock
\newblock
\shownote{\url{https://github.com/jupyter-server/jupyter_server/issues/521}}.


\bibitem[Jupyter(2021k)]%
        {notebook-pull-6160}
\bibfield{author}{\bibinfo{person}{Jupyter}.} \bibinfo{year}{2021}\natexlab{k}.
\newblock \bibinfo{title}{Notebook pull\#6160}.
\newblock
\newblock
\newblock
\shownote{\url{https://github.com/jupyter/notebook/pull/6160}}.


\bibitem[Jupyter(2022a)]%
        {first-way-to-run-Jupyter}
\bibfield{author}{\bibinfo{person}{Jupyter}.} \bibinfo{year}{2022}\natexlab{a}.
\newblock \bibinfo{title}{Basic Steps to Run Jupyter}.
\newblock
\newblock
\newblock
\shownote{\url{https://docs.jupyter.org/en/latest/running.html\#basic-steps}}.


\bibitem[Jupyter(2022b)]%
        {jupyterserver-post-information}
\bibfield{author}{\bibinfo{person}{Jupyter}.} \bibinfo{year}{2022}\natexlab{b}.
\newblock \bibinfo{title}{Detailed POST Inofrmation of Jupyterserver}.
\newblock
\newblock
\newblock
\shownote{\url{https://petstore.swagger.io/?url=https://raw.githubusercontent.com/jupyter/jupyter_server/master/jupyter_server/services/api/api.yaml\#/sessions/post_api_sessions}}.


\bibitem[Jupyter(2022c)]%
        {Jupyter}
\bibfield{author}{\bibinfo{person}{Jupyter}.} \bibinfo{year}{2022}\natexlab{c}.
\newblock \bibinfo{title}{Jupyter}.
\newblock
\newblock
\newblock
\shownote{\url{https://jupyter.org/}}.


\bibitem[Jupyter(2022d)]%
        {Notebook}
\bibfield{author}{\bibinfo{person}{Jupyter}.} \bibinfo{year}{2022}\natexlab{d}.
\newblock \bibinfo{title}{Jupyter Notebook}.
\newblock
\newblock
\newblock
\shownote{\url{https://github.com/jupyter/notebook}}.


\bibitem[Jupyter(2022e)]%
        {jupyterclient}
\bibfield{author}{\bibinfo{person}{Jupyter}.} \bibinfo{year}{2022}\natexlab{e}.
\newblock \bibinfo{title}{jupyterclient}.
\newblock
\newblock
\newblock
\shownote{\url{https://github.com/jupyter/jupyter_client}}.


\bibitem[Jupyter(2022f)]%
        {jupytercore}
\bibfield{author}{\bibinfo{person}{Jupyter}.} \bibinfo{year}{2022}\natexlab{f}.
\newblock \bibinfo{title}{jupytercore}.
\newblock
\newblock
\newblock
\shownote{\url{https://github.com/jupyter/jupyter_core}}.


\bibitem[Jupyter(2022g)]%
        {jupyterserver}
\bibfield{author}{\bibinfo{person}{Jupyter}.} \bibinfo{year}{2022}\natexlab{g}.
\newblock \bibinfo{title}{Jupyterserver}.
\newblock
\newblock
\newblock
\shownote{\url{https://github.com/jupyter-server/jupyter_server}}.


\bibitem[Jupyter(2022h)]%
        {jupyterserver-architecture}
\bibfield{author}{\bibinfo{person}{Jupyter}.} \bibinfo{year}{2022}\natexlab{h}.
\newblock \bibinfo{title}{Jupyterserver Architecture}.
\newblock
\newblock
\newblock
\shownote{\url{https://jupyter-server.readthedocs.io/en/latest/developers/architecture.html}}.


\bibitem[Jupyter(2022i)]%
        {jupyterserver-auto-launch}
\bibfield{author}{\bibinfo{person}{Jupyter}.} \bibinfo{year}{2022}\natexlab{i}.
\newblock \bibinfo{title}{Launching a bare Jupyterserver}.
\newblock
\newblock
\newblock
\shownote{\url{https://jupyter-server.readthedocs.io/en/latest/users/launching.html}}.


\bibitem[Jupyter(2022j)]%
        {jupyterserver-workflow}
\bibfield{author}{\bibinfo{person}{Jupyter}.} \bibinfo{year}{2022}\natexlab{j}.
\newblock \bibinfo{title}{Workflow in Jupyterserver}.
\newblock
\newblock
\newblock
\shownote{\url{https://jupyter-server.readthedocs.io/en/latest/developers/architecture.html\#create-session-workflow}}.


\bibitem[Kang et~al\mbox{.}(2021)]%
        {Kang:2021}
\bibfield{author}{\bibinfo{person}{DaYe Kang}, \bibinfo{person}{Tony Ho},
  \bibinfo{person}{Nicolai Marquardt}, \bibinfo{person}{Bilge Mutlu}, {and}
  \bibinfo{person}{Andrea Bianchi}.} \bibinfo{year}{2021}\natexlab{}.
\newblock \showarticletitle{ToonNote: Improving Communication in Computational
  Notebooks Using Interactive Data Comics}. In
  \bibinfo{booktitle}{\emph{Proceedings CHI}}.
\newblock


\bibitem[Kery and Myers(2018)]%
        {Kery:2018}
\bibfield{author}{\bibinfo{person}{Mary~Beth Kery} {and}
  \bibinfo{person}{Brad~A. Myers}.} \bibinfo{year}{2018}\natexlab{}.
\newblock \showarticletitle{Interactions for Untangling Messy History in a
  Computational Notebook}. In \bibinfo{booktitle}{\emph{Symposium on VL/HCC}}.
  \bibinfo{pages}{147--155}.
\newblock


\bibitem[Knuth(1984)]%
        {Knuth:1984}
\bibfield{author}{\bibinfo{person}{D.~E. Knuth}.}
  \bibinfo{year}{1984}\natexlab{}.
\newblock \showarticletitle{{Literate Programming}}.
\newblock \bibinfo{journal}{\emph{Comput. J.}} \bibinfo{volume}{27},
  \bibinfo{number}{2} (\bibinfo{year}{1984}), \bibinfo{pages}{97--111}.
\newblock


\bibitem[Koenzen et~al\mbox{.}(2020)]%
        {Koenzen:2020}
\bibfield{author}{\bibinfo{person}{Andreas~P. Koenzen},
  \bibinfo{person}{Neil~A. Ernst}, {and} \bibinfo{person}{Margaret-Anne~D.
  Storey}.} \bibinfo{year}{2020}\natexlab{}.
\newblock \showarticletitle{Code Duplication and Reuse in Jupyter Notebooks}.
  In \bibinfo{booktitle}{\emph{Symposium on VL/HCC}}. \bibinfo{pages}{1--9}.
\newblock


\bibitem[Koop and Patel(2017)]%
        {koop:2017}
\bibfield{author}{\bibinfo{person}{David Koop} {and} \bibinfo{person}{Jay
  Patel}.} \bibinfo{year}{2017}\natexlab{}.
\newblock \showarticletitle{Dataflow notebooks: encoding and tracking
  dependencies of cells}. In \bibinfo{booktitle}{\emph{9th USENIX Workshop on
  TaPP}}.
\newblock


\bibitem[Li et~al\mbox{.}(2021)]%
        {Li:2021}
\bibfield{author}{\bibinfo{person}{Xingjun Li}, \bibinfo{person}{Yuanxin Wang},
  \bibinfo{person}{Hong Wang}, \bibinfo{person}{Yang Wang}, {and}
  \bibinfo{person}{Jian Zhao}.} \bibinfo{year}{2021}\natexlab{}.
\newblock \showarticletitle{NBSearch: Semantic Search and Visual Exploration of
  Computational Notebooks}. In \bibinfo{booktitle}{\emph{Proceedings of CHI}}.
\newblock


\bibitem[Mathew and Stolee(2021)]%
        {Mathew:2021}
\bibfield{author}{\bibinfo{person}{George Mathew} {and}
  \bibinfo{person}{Kathryn~T. Stolee}.} \bibinfo{year}{2021}\natexlab{}.
\newblock \showarticletitle{Cross-language code search using static and dynamic
  analyses}. In \bibinfo{booktitle}{\emph{Proceedings of ESEC/FSE}}.
  \bibinfo{pages}{205--217}.
\newblock


\bibitem[nbconvert(2022)]%
        {nbconvert}
\bibfield{author}{\bibinfo{person}{nbconvert}.}
  \bibinfo{year}{2022}\natexlab{}.
\newblock \bibinfo{title}{nbconvert: Jupyter Notebook Conversion}.
\newblock
\newblock
\newblock
\shownote{\url{https://github.com/jupyter/nbconvert}}.


\bibitem[Nguyen et~al\mbox{.}(2018)]%
        {Nguyen:2018}
\bibfield{author}{\bibinfo{person}{Hai Nguyen}, \bibinfo{person}{David~A Case},
  {and} \bibinfo{person}{Alexander~S Rose}.} \bibinfo{year}{2018}\natexlab{}.
\newblock \showarticletitle{NGLview--interactive molecular graphics for Jupyter
  notebooks}.
\newblock \bibinfo{journal}{\emph{Bioinformatics}} \bibinfo{volume}{34},
  \bibinfo{number}{7} (\bibinfo{year}{2018}), \bibinfo{pages}{1241--1242}.
\newblock


\bibitem[Perez and Granger(2007)]%
        {Perez:2007}
\bibfield{author}{\bibinfo{person}{Fernando Perez} {and}
  \bibinfo{person}{Brian~E. Granger}.} \bibinfo{year}{2007}\natexlab{}.
\newblock \showarticletitle{IPython: A System for Interactive Scientific
  Computing}.
\newblock \bibinfo{journal}{\emph{Computing in Science Engineering}}
  \bibinfo{volume}{9} (\bibinfo{year}{2007}), \bibinfo{pages}{21--29}.
\newblock


\bibitem[Pimentel et~al\mbox{.}(2019)]%
        {Pimentel:2019}
\bibfield{author}{\bibinfo{person}{João~Felipe Pimentel},
  \bibinfo{person}{Leonardo Murta}, \bibinfo{person}{Vanessa Braganholo}, {and}
  \bibinfo{person}{Juliana Freire}.} \bibinfo{year}{2019}\natexlab{}.
\newblock \showarticletitle{A Large-Scale Study About Quality and
  Reproducibility of Jupyter Notebooks}. In
  \bibinfo{booktitle}{\emph{Proceedings of MSR}}. \bibinfo{pages}{507--517}.
\newblock


\bibitem[pydeps(2022)]%
        {pydeps}
\bibfield{author}{\bibinfo{person}{pydeps}.} \bibinfo{year}{2022}\natexlab{}.
\newblock \bibinfo{title}{pydeps}.
\newblock
\newblock
\newblock
\shownote{\url{https://pydeps.readthedocs.io/}}.


\bibitem[Python(2022)]%
        {re}
\bibfield{author}{\bibinfo{person}{Python}.} \bibinfo{year}{2022}\natexlab{}.
\newblock \bibinfo{title}{Python regex library}.
\newblock
\newblock
\newblock
\shownote{\url{https://docs.python.org/3/library/re.html}}.


\bibitem[Rehman(2019)]%
        {Rehman:2019}
\bibfield{author}{\bibinfo{person}{Mohammed~Suhail Rehman}.}
  \bibinfo{year}{2019}\natexlab{}.
\newblock \showarticletitle{Towards understanding data analysis workflows using
  a large notebook corpus}. In \bibinfo{booktitle}{\emph{Proceedings of ICMD}}.
  \bibinfo{pages}{1841--1843}.
\newblock


\bibitem[Romano et~al\mbox{.}(2021)]%
        {Romano:2021}
\bibfield{author}{\bibinfo{person}{Alan Romano}, \bibinfo{person}{Xinyue Liu},
  \bibinfo{person}{Yonghwi Kwon}, {and} \bibinfo{person}{Weihang Wang}.}
  \bibinfo{year}{2021}\natexlab{}.
\newblock \showarticletitle{An Empirical Study of Bugs in WebAssembly
  Compilers}. In \bibinfo{booktitle}{\emph{Proceedings of ASE}}.
  \bibinfo{pages}{42--54}.
\newblock


\bibitem[Rule et~al\mbox{.}(2018a)]%
        {Rule:2018}
\bibfield{author}{\bibinfo{person}{Adam Rule}, \bibinfo{person}{Ian Drosos},
  \bibinfo{person}{Aur{\'e}lien Tabard}, {and} \bibinfo{person}{James~D
  Hollan}.} \bibinfo{year}{2018}\natexlab{a}.
\newblock \showarticletitle{Aiding collaborative reuse of computational
  notebooks with annotated cell folding}.
\newblock \bibinfo{journal}{\emph{Proceedings of CHI}} (\bibinfo{year}{2018}),
  \bibinfo{pages}{1--12}.
\newblock


\bibitem[Rule et~al\mbox{.}(2018b)]%
        {rule2018}
\bibfield{author}{\bibinfo{person}{Adam Rule}, \bibinfo{person}{Aur{\'e}lien
  Tabard}, {and} \bibinfo{person}{James~D Hollan}.}
  \bibinfo{year}{2018}\natexlab{b}.
\newblock \showarticletitle{Exploration and explanation in computational
  notebooks}. In \bibinfo{booktitle}{\emph{Proceedings of CHI}}.
  \bibinfo{pages}{1--12}.
\newblock


\bibitem[Samuel and K{\"{o}}nig{-}Ries(2021)]%
        {Sheeba:2021}
\bibfield{author}{\bibinfo{person}{Sheeba Samuel} {and}
  \bibinfo{person}{Birgitta K{\"{o}}nig{-}Ries}.}
  \bibinfo{year}{2021}\natexlab{}.
\newblock \showarticletitle{ReproduceMeGit: {A} Visualization Tool for
  Analyzing Reproducibility of Jupyter Notebooks}. In
  \bibinfo{booktitle}{\emph{Proceedings of IPAW}},
  Vol.~\bibinfo{volume}{12839}. \bibinfo{pages}{201--206}.
\newblock


\bibitem[Seaman et~al\mbox{.}(2008)]%
        {Seaman:2008}
\bibfield{author}{\bibinfo{person}{Carolyn~B. Seaman}, \bibinfo{person}{Forrest
  Shull}, \bibinfo{person}{Myrna Regardie}, \bibinfo{person}{Denis Elbert},
  \bibinfo{person}{Raimund~L. Feldmann}, \bibinfo{person}{Yuepu Guo}, {and}
  \bibinfo{person}{Sally Godfrey}.} \bibinfo{year}{2008}\natexlab{}.
\newblock \showarticletitle{Defect Categorization: Making Use of a Decade of
  Widely Varying Historical Data}. In \bibinfo{booktitle}{\emph{Proceedings of
  ESEM}}. \bibinfo{pages}{149–157}.
\newblock


\bibitem[Shen(2014)]%
        {shen2014interactive}
\bibfield{author}{\bibinfo{person}{Helen Shen}.}
  \bibinfo{year}{2014}\natexlab{}.
\newblock \showarticletitle{Interactive notebooks: Sharing the code}.
\newblock \bibinfo{journal}{\emph{Nature}} \bibinfo{volume}{515},
  \bibinfo{number}{7525} (\bibinfo{year}{2014}), \bibinfo{pages}{151--152}.
\newblock


\bibitem[Shen et~al\mbox{.}(2021)]%
        {Shen:2021}
\bibfield{author}{\bibinfo{person}{Qingchao Shen}, \bibinfo{person}{Haoyang
  Ma}, \bibinfo{person}{Junjie Chen}, \bibinfo{person}{Yongqiang Tian},
  \bibinfo{person}{Shing{-}Chi Cheung}, {and} \bibinfo{person}{Xiang Chen}.}
  \bibinfo{year}{2021}\natexlab{}.
\newblock \showarticletitle{A comprehensive study of deep learning compiler
  bugs}. In \bibinfo{booktitle}{\emph{Proceedings of ESEC/FSE}},
  \bibfield{editor}{\bibinfo{person}{Diomidis Spinellis},
  \bibinfo{person}{Georgios Gousios}, \bibinfo{person}{Marsha Chechik}, {and}
  \bibinfo{person}{Massimiliano~Di Penta}} (Eds.). \bibinfo{pages}{968--980}.
\newblock


\bibitem[Thung et~al\mbox{.}(2012)]%
        {Thung:2012}
\bibfield{author}{\bibinfo{person}{Ferdian Thung}, \bibinfo{person}{Shaowei
  Wang}, \bibinfo{person}{David Lo}, {and} \bibinfo{person}{Lingxiao Jiang}.}
  \bibinfo{year}{2012}\natexlab{}.
\newblock \showarticletitle{An Empirical Study of Bugs in Machine Learning
  Systems}. In \bibinfo{booktitle}{\emph{Proceedings of ISSRE}}.
  \bibinfo{pages}{271--280}.
\newblock


\bibitem[Tornado(2022)]%
        {tornado}
\bibfield{author}{\bibinfo{person}{Tornado}.} \bibinfo{year}{2022}\natexlab{}.
\newblock \bibinfo{title}{Tornado: a Python web framework}.
\newblock
\newblock
\newblock
\shownote{\url{https://www.tornadoweb.org/en/stable/}}.


\bibitem[Traitlets(2021)]%
        {Traitlets}
\bibfield{author}{\bibinfo{person}{Traitlets}.}
  \bibinfo{year}{2021}\natexlab{}.
\newblock \bibinfo{title}{Traitlets framework}.
\newblock
\newblock
\newblock
\shownote{\url{https://github.com/ipython/traitlets}}.


\bibitem[Vasilescu et~al\mbox{.}(2015)]%
        {Vasilescu:2015}
\bibfield{author}{\bibinfo{person}{Bogdan Vasilescu}, \bibinfo{person}{Yue Yu},
  \bibinfo{person}{Huaimin Wang}, \bibinfo{person}{Premkumar Devanbu}, {and}
  \bibinfo{person}{Vladimir Filkov}.} \bibinfo{year}{2015}\natexlab{}.
\newblock \showarticletitle{Quality and Productivity Outcomes Relating to
  Continuous Integration in GitHub}. In \bibinfo{booktitle}{\emph{Proceedings
  of ESEC/FSE}}. \bibinfo{pages}{805–816}.
\newblock


\bibitem[Wang et~al\mbox{.}(2021b)]%
        {Wang:2021An}
\bibfield{author}{\bibinfo{person}{Dinghua Wang}, \bibinfo{person}{Shuqing Li},
  \bibinfo{person}{Guanping Xiao}, \bibinfo{person}{Yepang Liu}, {and}
  \bibinfo{person}{Yulei Sui}.} \bibinfo{year}{2021}\natexlab{b}.
\newblock \showarticletitle{An exploratory study of autopilot software bugs in
  unmanned aerial vehicles}. In \bibinfo{booktitle}{\emph{Proceedings of
  ESEC/FSE}}. \bibinfo{pages}{20--31}.
\newblock


\bibitem[Wang et~al\mbox{.}(2020a)]%
        {Wang:2020ASE}
\bibfield{author}{\bibinfo{person}{Jiawei Wang}, \bibinfo{person}{Tzu-yang
  Kuo}, \bibinfo{person}{Li Li}, {and} \bibinfo{person}{Andreas Zeller}.}
  \bibinfo{year}{2020}\natexlab{a}.
\newblock \showarticletitle{Assessing and Restoring Reproducibility of Jupyter
  Notebooks}. In \bibinfo{booktitle}{\emph{Proceedings of ASE}}.
  \bibinfo{pages}{138–149}.
\newblock


\bibitem[Wang et~al\mbox{.}(2020b)]%
        {Wang:2020}
\bibfield{author}{\bibinfo{person}{Jiawei Wang}, \bibinfo{person}{Li Li}, {and}
  \bibinfo{person}{Andreas Zeller}.} \bibinfo{year}{2020}\natexlab{b}.
\newblock \showarticletitle{Better Code, Better Sharing: On the Need of
  Analyzing Jupyter Notebooks}. In \bibinfo{booktitle}{\emph{Proceedings of the
  ICSE-NIER}}. \bibinfo{pages}{53–56}.
\newblock


\bibitem[Wang et~al\mbox{.}(2021a)]%
        {Wang:2021}
\bibfield{author}{\bibinfo{person}{Jiawei Wang}, \bibinfo{person}{Li Li}, {and}
  \bibinfo{person}{Andreas Zeller}.} \bibinfo{year}{2021}\natexlab{a}.
\newblock \showarticletitle{Restoring Execution Environments of Jupyter
  Notebooks}. In \bibinfo{booktitle}{\emph{Proceedings of ICSE}}.
  \bibinfo{pages}{1622--1633}.
\newblock


\bibitem[webpack(2022)]%
        {webpack}
\bibfield{author}{\bibinfo{person}{webpack}.} \bibinfo{year}{2022}\natexlab{}.
\newblock \bibinfo{title}{Webpack: a module bundler for JavaScript}.
\newblock
\newblock
\newblock
\shownote{\url{https://webpack.js.org/}}.


\bibitem[Weinman et~al\mbox{.}(2021)]%
        {Weinman:2021}
\bibfield{author}{\bibinfo{person}{Nathaniel Weinman},
  \bibinfo{person}{Steven~M Drucker}, \bibinfo{person}{Titus Barik}, {and}
  \bibinfo{person}{Robert DeLine}.} \bibinfo{year}{2021}\natexlab{}.
\newblock \showarticletitle{Fork It: Supporting stateful alternatives in
  computational notebooks}. In \bibinfo{booktitle}{\emph{Proceedings of CHI}}.
  \bibinfo{pages}{1--12}.
\newblock


\bibitem[ZeroMQ(2022)]%
        {ZeroMQ}
\bibfield{author}{\bibinfo{person}{ZeroMQ}.} \bibinfo{year}{2022}\natexlab{}.
\newblock \bibinfo{title}{ZeroMQ}.
\newblock
\newblock
\newblock
\shownote{\url{https://zeromq.org/}}.


\bibitem[Zhang et~al\mbox{.}(2018)]%
        {Zhang:2018}
\bibfield{author}{\bibinfo{person}{Yuhao Zhang}, \bibinfo{person}{Yifan Chen},
  \bibinfo{person}{Shing-Chi Cheung}, \bibinfo{person}{Yingfei Xiong}, {and}
  \bibinfo{person}{Lu Zhang}.} \bibinfo{year}{2018}\natexlab{}.
\newblock \showarticletitle{An Empirical Study on TensorFlow Program Bugs}. In
  \bibinfo{booktitle}{\emph{Proceedings of ISSTA}}. \bibinfo{pages}{129–140}.
\newblock


\bibitem[Zhu et~al\mbox{.}(2021)]%
        {Zhu:2021}
\bibfield{author}{\bibinfo{person}{Chenguang Zhu}, \bibinfo{person}{Ripon~K.
  Saha}, \bibinfo{person}{Mukul~R. Prasad}, {and} \bibinfo{person}{Sarfraz
  Khurshid}.} \bibinfo{year}{2021}\natexlab{}.
\newblock \showarticletitle{Restoring the Executability of Jupyter Notebooks by
  Automatic Upgrade of Deprecated APIs}. In
  \bibinfo{booktitle}{\emph{Proceedings of ASE}}. \bibinfo{pages}{240--252}.
\newblock


\end{thebibliography}

\end{document}